\documentclass[preprint,10pt]{aastex}   
\usepackage{natbib}
\newcommand{\hi}{{H$\,$\footnotesize I}}
\newcommand{\sm}{$\sim\,$}
\newcommand{\bb}{$B$}

\newcommand{\kms}{\,km\,s$^{-1}$}
\newcommand{\df}{\mbox{DEF}}
\newcommand{\dfbis}{<DEF>}
\newcommand{\fdf}{$F_\mathrm{DEF}$}
\newcommand{\fdef}{$F_\mathrm{DEF}^\mathrm{c}$}

\newcommand{\mtc}{$M_{B_\mathrm{T}}^\mathrm{c}$}
\newcommand{\wc}{$W_\mathrm{20}^\mathrm{c}$}
\newcommand{\vel}{$v_\mathrm{sys}$}
\newcommand{\subhi}{_\mathrm{H\mbox{\tiny I}}}
\newcommand{\mhi}{M\subhi}
\newcommand{\shi}{\overline{\Sigma}\subhi}
\newcommand{\fhi}{F\subhi}
\newcommand{\aas}{A\&AS}       
\newcommand{\sci}{Science}     

\slugcomment{Accepted for publication by The Astronomical Journal}
\shorttitle{The 3D Structure of Virgo}
\shortauthors{Solanes et al.}

\begin{document}

\title{The 3D Structure of the Virgo Cluster Region from Tully-Fisher
and \hi\ Data}


\author{Jos\'e M.\ Solanes} 
\affil{Departament d'Enginyeria Inform\`atica i
Ma\-te\-m\`a\-ti\-ques. Campus Sescelades, Universitat Rovira i
Virgili. Avda.\ Pa\"\i sos Catalans, 26; E--43007~Ta\-rra\-go\-na,
Spain}
\author{Teresa Sanchis and Eduard Salvador-Sol\'e}
\affil{Departament d'Astronomia i Meteorologia and CER d'Astrof\'\i
sica, F\'\i sica de Part\'\i cules i Cosmologia, Universitat de
Barcelona. Av.\ Diagonal 647; E--08028~Barcelona, Spain}
\and
\author{Riccardo Giovanelli and Martha P.\ Haynes} 
\affil{Center for Radiophysics and Space Research and National
Astronomy and Ionosphere Center\footnote{The National Astronomy and
Ionosphere Center is operated by Cornell University under a cooperative
agreement with the National Science Foundation.},\\ Cornell University;
Ithaca, NY 14853}
\email{jsolanes@etse.urv.es, (tsanchis,eduard)@am.ub.es,
(riccardo,haynes)@astro.cornell.edu}

\begin{abstract}
The distances and \hi\ contents of 161 spiral galaxies in the region of
Virgo cluster are used to gain insight into the complicated structure
of this galaxy system. Special attention has been paid to the
investigation of the suggestion presented in an earlier work that some
peripheral Virgo groups may contain strongly gas-deficient spirals. 

The three-dimensional galaxy distribution has been inferred from
quality distance estimates obtained by averaging distance moduli based
upon the Tully-Fisher relationship taken from eight published datasets
previously homogenized, resulting in a relation with a dispersion of
0.41~mag. Previous findings that the spiral distribution is
substantially more elongated along the line-of-sight than in the plane
of the sky are confirmed by the current data. In addition, an important
east-west disparity in this effect has been detected. The overall
width-to-depth ratio of the Virgo cluster region is about 1\,:\,4, with
the most distant objects concentrated in the western half. The
filamentary structure of the spiral population and its orientation are
also reflected by the \hi-deficient objects alone. The \hi\ deficiency
pattern shows a central enhancement extending from \sm16 to 22 Mpc in
line-of-sight distance; most of this enhancement arises from galaxies
that belong to the Virgo cluster proper. However, significant gas
deficiencies are also detected outside the main body of the cluster in
a probable group of galaxies at line-of-sight distances \sm25--30 Mpc,
lying in the region dominated by the southern edge of the M49
subcluster and clouds W$^\prime$ and W, as well as in various
foreground galaxies. In the Virgo region, the \hi\ content of the
galaxies then is not a straightforward indicator of cluster membership.
\end{abstract}


\section{Introduction}

This article is the continuation of a series of papers on the \hi\
content of spirals from the 21 cm line of neutral hydrogen data devoted
to examine the extent to which the cluster environment influences the
evolution of the galaxies.

The starting point, developed in \citet*[hereafter Paper~I]{SGH96}, was
the establishment of reliable standards of \hi\ mass for the various
morphological subgroups of luminous spirals from a complete
\hi-flux-limited sample of these galaxy types in low density
environments. That work was followed by a second paper examining the
possible connections between gas deficiency and the properties of both
the underlying galaxies and their environment in the fields of eighteen
nearby clusters by \citet*[hereafter Paper~II]{Sol01}. The main
motivation was to gain insight into the mechanisms responsible for the
atomic gas depletion. While no clearly discriminating circumstances
were found among those clusters which show significant \hi\ deficiency
and those which do not, this work definitely confirmed previous
findings \citep[e.g.,][]{GH85,HG86,Mag88} that in \hi-deficient
clusters the proportion of gas-poor spirals increases monotonically
towards the center. Moreover, \citeauthor{Sol01} clearly demonstrated,
as first suggested by \citet*{Dre86}, that \hi-deficient objects move
on orbits more radial than those of their gas-rich counterparts. This
result made a strong case for the ram-pressure stripping of the spirals
by the hot X-ray emitting intracluster medium (ICM) as the most likely
process responsible for the gas deficiencies observed in rich cluster
environments.

The wealth of 21-cm data gathered for the Virgo region in
\citeauthor{Sol01} also made it possible to examine the distribution in
two-dimensional space of the neutral gas deficiency in the Virgo
central area. The sky distribution of \hi\ deficiency was found to be
in overall agreement with the radial pattern characteristic of rich
clusters, showing that the maximum depletion occurred at the cluster
center. But quite unexpectedly, the same map of the HI deficiency
pattern, a variation of which is produced here as Fig.~\ref{virgohiam},
also revealed peripheral groups of galaxies with a dearth of atomic
hydrogen but found in areas where the density of X-ray luminous gas is
very low, raising into question the feasibility that the ram-pressure
of the ICM was responsible for the observed \hi\ deficiency. 

In the current paper, we conduct a further investigation into the
nature of and conditions within the three-dimensional structure of the
Virgo region, by incorporating into the analysis the \hi\ content of
its spiral population. The proximity of the region under study
facilitates the gathering of a large number of 21-cm single-dish
observations, which we complement with a large number of Tully-Fisher
\citetext{\citeyear{TF77}, hereafter TF} distance estimates also
reported in the literature. Section~\ref{data} presents a catalog of
161 galaxies with good \hi\ and TF distance measurements. After
reviewing in Section~\ref{hidef} the manner in which the \hi\
deficiency is calculated, our galaxy sample is used in the following
two sections to discuss, first the radial pattern of \hi\ deficiency,
and then its spatial distribution in the Virgo region. We conclude with
a summary and some remarks in Section~\ref{conclusions}.

\section{Observational Data}\label{data}
\subsection{Galaxy Selection}

The backbone of the present study is the complete spiral sample taken
from the \emph{Virgo Cluster Catalog} \citep*[hereafter VCC]{VCC} by
\citet*[YFO97]{YFO97} to study the Virgo cluster using the \bb-band TF
relation.  The \citeauthor{YFO97} sample has been supplemented by data
from seven other studies of the Virgo cluster, likewise presenting TF
distance estimates at various wavelengths. Table~\ref{datasets} lists
the different sources of TF distances for Virgo objects included in the
present work, along with the number of galaxies included in our
catalog. These datasets include virtually all spiral galaxies used to
date in the application of the TF relation to study the Virgo region.

For the current purpose, we have selected from the original
catalogs listed in Table~\ref{datasets} only galaxies with heliocentric
radial velocities below the well-defined gap near 3000 \kms\ that
neatly isolates the Virgo region in redshift space \citep*{BPT93}. In
addition, we have focused on galaxies located in the region bounded by
$\rm 12^h\le R.A.\le 13^h$ and $\rm 0\degr\le Decl.\le +25\degr$
(throughout the paper equatorial coordinates are referred to the
B1950.0 equinox), which encompasses the VCC survey boundary and is
centered on the classical Virgo~I cluster \citep{dVau61}. Thus,
wherever we use the terminology ``Virgo cluster region'', it should be
kept in mind that objects located in the ``Virgo Southern Extension''
\citep{Tul82} or Virgo~II cluster, i.e., with $\rm Decl.<+5\degr$, are
indeed largely excluded. Note that our selection procedure also implies
that all the galaxies included in our sample (even those classified as
background objects in VCC) are expected to have peculiar motions
influenced by the central mass concentration of the cluster.

Our initial selection of Virgo galaxies includes a total of 198
objects, which we summarize in Table~\ref{distances}. This table
contains, among other information, the distance moduli to each galaxy
given in the eight TF studies on which our investigation is based. For
seven of these galaxies, we also list their Cepheid distances given in
the final results from the Hubble Space Telescope Key Project to
measure the Hubble constant by \citet{Fre01} corrected for the effects
of metallicity.

\subsection{Homogenization of the distances to individual galaxies}
\label{homogen} 

Since our compilation of TF distances was built from eight datasets
that contain sometimes inconsistent data, the average of the
measurements available for each object does not necessarily provide the
best estimate of the galaxy distances. We thus have attempted first to
reduce the data to a homogeneous system by eliminating systematic
differences among the different sources. We have carried out this task
by means of a recursive procedure applied separately to each of the
eight Virgo datasets listed in Table~\ref{datasets} and composed of the
following steps:

(i) For each dataset, we take only those galaxies which also have
   distance measurements in any of the other seven datasets and
   calculate for each of those objects the arithmetic mean of all their
   distance estimates, as well as the standard deviation of the
   individual measurements.

(ii) Then, using only the galaxies in the chosen catalog which have
   multiple distance measurements, we determine the linear regression
   of the individual distances against the average values, applying a
   $3\sigma$ clipping to remove those points which deviate
   significantly from the regression line. The regression coefficients
   are then recalculated and the $3\sigma$ rejection procedure is
   repeated until no more galaxies are rejected. Once a datum is
   flagged as highly deviant, it retains this status for the rest of
   the procedure.

(iii) The regression coefficients calculated at the end of step (ii)
   are used to transform all distance measurements of the dataset
   under scrutiny, including those not listed elsewhere, into the
   system of mean distances defined by the sample members with multiple
   observations.

The standardization to the mean values affects each dataset
differently. So once steps (i), (ii), and (iii) have been performed for
the eight source catalogs (excluding the data already flagged as most
deviant), the procedure is repeated again until all the corrections
become negligible. Convergence to a homogeneous system of mean
distances is achieved in only a few cycles.

When \emph{all} the distance measurements available for an individual
object show residuals inconsistent with the regression lines, we
ascribe this to a possible source misidentification and accordingly
exclude the galaxy from the calculations. This seems to be the case of
the galaxies V0975 and V1678, with only two distance measurements each
that disagree beyond allowed uncertainty. We have also excluded from
the homogenization process the seven galaxies with Cepheid distances:
V0596, V1375, V1555, V1562, V1615, V1943, and N4725. For these galaxies
we use the Cepheid measurements and their quoted errors.

The distances obtained with the procedure described above and their
associated $1\sigma$ uncertainties are listed in the last column of
Table~\ref{distances}. The rms error of the individual distance moduli
is $0.29$~mag, which translates to an uncertainty of about 13\% in the
redshift-independent distance measurements. This means that for an
individual galaxy at, say, 20 Mpc, we expect a typical error in the
distance of \sm2.5 Mpc. This implies a sufficient level of accuracy in
the measurement of individual distances to allow us to explore the
gross features of the three-dimensional structure of the Virgo cluster.

\subsection{The 21-cm sample}\label{21cm}

For the homogenization of the distance moduli, we have taken all the
galaxies listed in two or more of the source TF catalogs (i.e., with
multiple distance measures), regardless of their Hubble type, in order
to minimize the impact of statistical ``shot noise''. Since irregular
and bulge-dominated galaxies may give, however, unreliable TF distances
and/or \hi\ content measurements, we re-examined the morphologies of
the galaxies and picked up only those with Hubble types ranging from
$T=1$ (Sa) to $T=9$ (Sm), as given in the \emph{Third Reference
Catalogue of Bright Galaxies}
\citep*[hereafter RC3]{RC3}.

The principal source for the \hi\ line fluxes is the \emph{Arecibo
General Catalog} (hereafter AGC), a private database maintained by RG
and MPH, which contains an extensive compilation of 21-cm-line
measurements collected from a large number of sources. Among the
galaxies with a clear spiral morphology, there are 15 for which the AGC
did not provide useful data, including 2 galaxies that overlap with
obvious optical companions and 3 non-detections. We found \hi\ flux
measures for 9 of these objects in \emph{A General Catalog of \hi\
Observations of Galaxies} \citep{HR89} and in the \emph{Lyon-Meudon
Extragalactic Database} (LEDA). All the observational values have been
corrected for the effects of random pointing errors, source extent, and
internal \hi\ absorption following \citet{HG84}, except the nine
non-AGC fluxes and the only non-detection, V0522, which have been
corrected only for internal \hi\ self-absorption and assigned an 'H'
flag in the final dataset.

The AGC is also adopted as the source of other observational parameters
required by our study, such as the equatorial coordinates of the
galaxies, their visual optical diameters, which are involved in the
determination of the \hi\ deficiency (see next section), and their
heliocentric radial velocities, which we transform to systemic
velocities, \vel, by referring them to the kinematic frame of the Local
Group, taken equal to 308 \kms\ towards $(l,b) = (105\degr,-7\degr)$
\citep*{YTS77}.

On the other hand, for the \hi-line width, a parameter of lesser
importance in this investigation, the heterogeneity of the measures
contained in the AGC prompted us to adopt instead the
inclination-corrected values of the line width at 20\% level of the
line-profile peak, \wc, listed in \citeauthor{YFO97}. For most of the
galaxies that concern us here, these authors provide a set of
observations standardized into line widths measured at the Arecibo
circular feed following a similar process to that carried out here with
the distance moduli. For galaxies not included in the
\citeauthor{YFO97} sample, we use the values of $W_\mathrm{20}$ and
inclination quoted in LEDA, except for the galaxy V1043, not listed in
either of these two catalogs, for which we adopt the corresponding
measurements by \citeauthor{MAH80} (measurements other than from
\citeauthor{YFO97} are flagged 'W' in the final dataset). Furthermore,
we have excluded spiral galaxies with \wc$\;\le 100\,$\kms\ to reduce
the error induced from turbulent disk motion. We do not impose an
inclination cut because this parameter does not play any explicit role
in our investigation (remember that, for most of the galaxies in our
dataset, we are adopting values of the line width already corrected
for inclination). Nevertheless, objects with $i<45\degr$ in LEDA are
warned with an 'i' flag in our final catalog.

After all these selections, we end up with a sample of 161 spiral
galaxies with reliable \hi\ content and distance data, hereafter called
the ``21-cm sample'', which is used below to assess the spatial
distribution of the neutral gas deficiency in the Virgo cluster
region. As stated in \S~\ref{bias}, the scatter of the best fitting TF
template for the 161 galaxies is 0.41~mag. Hence, the uncertainty in
the distance modulus of the individual galaxies in this dataset is
comparable to the scatter of the most accurate TF template relations
currently available.

All the parameters relevant to our investigation are listed in
Table~\ref{parameters}, where we have included two different
measurements of the \hi\ deficiency and the absolute \bb-magnitude of
the galaxies, \mtc, calculated from their total apparent corrected
\bb-magnitude listed in LEDA and our distance estimate. The sky
distribution of the members of the 21-cm sample is presented in
Figure~\ref{virgohiam}.

\section{The Diagnosis of \hi\ Deficiency}\label{hidef} 

\hi\ deficiency is often quantified by the parameter $\dfbis$ defined as
\begin{equation}\label{def2}
\dfbis\;=\;\langle\log\mhi(D_\mathrm{opt},T)\rangle-\log\mhi\;,
\end{equation} 
\citetext{e.g., \citealt*{CBG80}; \citealt{HG84}; \citeauthor{SGH96}},
where $\mhi$ is the \hi\ mass of the galaxy in solar units, and the
angular brackets on the right of the equal sign indicate the expected
value of this quantity inferred from a sample of field galaxies of the
same \emph{optical} linear diameter $D_\mathrm{opt}$ and morphological
type $T$. The most recent determinations of the expectation values for
the \hi\ mass as a function of the size and morphology of the galaxies
are given in \citeauthor{SGH96} in the form of linear regressions that
imply power law relationships of the type $\mhi\propto
D_\mathrm{opt}^n$, with the values of $n$ oscillating between about 1.7
for Sc's and 1.2 for earlier spiral types.

For the present study, however, we need to use a calibrator for the
neutral gas deficiency not tied to the distance to the galaxies. Given
that the $\mhi-D_\mathrm{opt}$ relationships do not deviate
substantially from a constant \hi\ surface density, especially for the
latest spiral types, it is reasonable to adopt the distance-independent
approximation to equation~(\ref{def2}) based on the difference of the
logarithms of the expected and observed values of this latter quantity
\begin{equation}\label{def1} 
\df=\langle\log\shi (T)\rangle-\log\shi\;,
\end{equation}
where $\shi$ is the mean \emph{hybrid} \hi\ surface density, which can
be calculated directly from the ratio of the observables $\fhi$ and the
apparent optical diameter of the galaxy, $a_\mathrm{opt}^2$, given in
arcmin \citetext{see also \citeauthor{SGH96}}. The adopted values for
$\langle\log\shi (T)\rangle$ are: 0.24 units for Sa, Sab; 0.38 for Sb;
0.40 for Sbc; 0.34 for Sc; and 0.42 for later spiral types. 

The values of \hi\ deficiency for the galaxies of the 21-cm sample
calculated using the two definitions given above are listed in columns
(4) and (5) of Table~\ref{parameters} ---to gauge the numerical
values, they can be compared with the overall scatter of 0.24 units
shown by field galaxies \citetext{\citeauthor{SGH96}}. A brief glance
at the figures shows that the two measurements are indeed very close
for most galaxies. Hence, statistical measurements of the \hi\
deficiency that rely on values of $\df$ taken from subsets of objects
not segregated according to size should give unbiased estimates of this
property. We note also that the error in $\dfbis$ resulting from the
propagation of the uncertainty in the distance modulus (see
\S~\ref{homogen}) is equal to 0.079 units for an Sb galaxy, while the
model scatter is equal to 0.126 units. This means that a substantial
portion of the error budget of this model can be attributed exclusively
to the uncertainty of the galactic distances.

\section{Possible Bias of \hi\ Deficiency in the Distance}\label{bias}

A potential problem with the determination of TF distances from \hi\
line widths, first brought into consideration by \citet{Guh88} and
\citet{Tee92}, results from the possibility that gas stripping can
reduce the size of the \hi\ disks to radii smaller than the turnover of
the rotation curve \citep[see, e.g., ][]{Cay94}. This would produce
objects that are too luminous for their line width, therefore resulting
in TF distances that are artificially underestimated. Possible
evidence of this effect has been cited in the Virgo cluster studies by
\citet{FOY93}, \citeauthor{YFO97}, and \citeauthor{FTS98}, though no
environmental dependences on TF distance determinations have been found
in more distant clusters \citep*{Gio97,Dal01} using combined \hi\ and
optical rotation width datasets.

We have investigated the possible alteration of distance resulting from
atomic gas deficiency in our Virgo data by inspecting the positions of
the members of the 21-cm sample on the \mtc-$\log$\wc\ plane (i.e., the
TF relation) according to their \hi\ deficiency. The two linear
regressions of the data excluding \hi-deficient objects (i.e., those
with $\df\ge 2\sigma$) are: \mtc$\;=-6.75\log$\wc$-2.55$ and
$\log$\wc$\;=-0.13$\mtc$-0.05$. More importantly, the number of
gas-poor galaxies on each side of the regression lines is about the
same. Similar behaviors are obtained when the threshold for the
exclusion of the \hi-deficient objects is increased up to 3 and
$4\sigma$ (the latter value implying roughly a factor 10 decrement in
the \hi\ mass), with regression lines obeying equations almost
identical to the former ones. These results are consistent with a
scenario in which the gas deficiency does not have any noticeable
effect on the distance estimates, even for the objects most severely
depleted in their interstellar \hi\ gas. This conclusion is also
supported by the observation that for the data points corresponding to
the most \hi-deficient galaxies the values of $\dfbis$ do not appear to
be systematically lower than those calculated from the
distance-independent parameter $\df$.

The absence of systematic deviations from the mean values of the TF
relation with increasing gas deficiency suggests that, in any event,
the value of the distance underestimates that could affect our most
\hi-deficient galaxies should be commonly smaller than the rms residual
about the regression line, which has a value of 0.41 mag (i.e.,
should be lower than 19\%). We note that this scatter is only
slightly larger than the values around 0.35 mag found in TF templates
based on the combination of measurements in many clusters
\citetext{e.g., \citeauthor{Gav99}, \citealt{Dal99}} and fully
comparable to the intrinsic scatter of 0.43 mag found by \citet{Sak00}
for nearby galaxies with Cepheid distances.

\section{The \hi\ Deficiency in Virgo}\label{hiversusd}

Numerous studies \citep[e.g.,][]{GH85,HG86,Mag88,Cay94,Bra00} reveal
that gas-poor galaxies tend to be more abundant in the centers of rich
galaxy clusters than in their periphery. Virgo is less rich and younger
than the classical Abell clusters and is characterized by a lower X-ray
luminosity and larger spiral fraction than Coma-like clusters. Probably
as a result, although it does contain a substantial fraction of
\hi-deficient galaxies \citetext{e.g., \citealp{HG86}; 
\citeauthor{Sol01}; see also below}, the degree of \hi\ deficiency 
is not observed to increase towards the center as dramatically as in
other rich clusters. However, because of its proximity, even
strongly gas-poor galaxies remain detected, allowing precise
determination of higher degrees of the \hi-deficiency, whereas in more
distant clusters only lower limits to this parameter can be derived.

The contour map of \hi\ deficiency shown in Figure~\ref{virgohiam}
illustrates that the maximum of the gas deficiency distribution
coincides with the position of the central cD galaxy, M87, where the
projected galaxy and intracluster gas densities are also the
highest. But this map also reveals other zones of significant
deficiency at sky positions dominated by background subclumps which lie
at substantial clustercentric distances.

\subsection{Radial Pattern}

The first lines of evidence that there is an excess of highly deficient
galaxies on the outskirts of the Virgo cluster are presented here by
means of Figures~\ref{hivsd} and \ref{hivsd3}. Figure~\ref{hivsd} shows
the values of $\df$ for the 161 members of the 21-cm sample as a
function of their line-of-sight (LOS) distance, $d$. This diagram
illustrates that most of the galaxies with substantial deficiencies in
the Virgo cluster region are localized in a broad range of projected
distances, which stretches from about 10 to 30 Mpc along the LOS. A few
more gas deficient objects lie beyond 40 Mpc.

By transforming the sky positions of the galaxies and their LOS
distances to rectangular coordinates, we can also inspect the behavior
of the \hi\ deficiency as a function of the \emph{three-dimensional}
radial distance, $r$, from the center of Virgo. We adopt the standard
identification of the cluster center at the position of M87, given by
the sky coordinates $(12^{\rm h}28\fm3,12\degr 40\arcmin)$ and a
distance modulus of 31.11 mag quoted in LEDA, which translates to a LOS
distance of 16.7 Mpc. The results are shown in Figure~\ref{hivsd3},
where we adopt two different representations of the radial run of the
\hi\ deficiency: one based on the parameter \fdf\ used in
\citeauthor{Sol01}, which measures the relative populations of
deficient and normal spirals, and the other based directly on the
averaged values of \df. In both cases the data have been binned into
annuli containing 16 galaxies per ring, with the final bin having 17,
in order to increase the statistical weight of the scarcer low- and
high-distance objects.

We see that, for $r\lesssim 4$ Mpc, the radial behavior of the gas
deficiency is consistent with the pattern exhibited by the composite
sample of 11 \hi-deficient clusters investigated previously in
\citeauthor{Sol01} (see Figure~4 therein): it decreases almost
monotonically towards normalcy with increasing distance from the
cluster center. But at greater Virgocentric distances this tendency is
broken by a series of secondary maxima, more conspicuous in the radial
run of \fdf\ because of its higher sensitivity to localized
enhancements, caused by regional enhancements of gas deficiency where
this parameter reaches values several times larger than ---and clearly
inconsistent with--- the field expectation (\sm2--3\%). Besides, as
illustrated in Figure~\ref{hivsd}, the spirals on the Virgo cluster
outskirts are not only more likely to be deficient in \hi\ than field
objects, but they also reach gas deficiencies typical of the cluster
core. The reader, however, should be aware of the fact that this same
sort of careful analysis of \hi\ deficiency at large clustercentric
distances has not been performed on other clusters. So, it is not
unfeasible that the differences in the radial pattern can be explained
simply by the bias that arises from Virgo's proximity which leads to
(a) much larger number of 21-cm observations, (b) more stringent values
of $\df$, and (c) more accurate TF distance estimates.

Further inspection of Figure~\ref{hivsd} (see also \S~\ref{spherical})
reveals that the central peak in the radial pattern of the \hi\
deficiency is the result of the accumulation of highly deficient
galaxies in the interval of LOS distances ranging from \sm16 up to 22
Mpc. This range coincides essentially with the distribution of the
bright ellipticals associated with the cluster core \citep{NT00}. The
second local maxima visible in the radial run of \fdf\ is produced by
galaxies with extreme deficiencies at $d\lesssim 15$ Mpc, while the
peak most distant from the cluster core obeys to several gas-deficient
objects at LOS distances between about 25 and 30 Mpc. As we show in the
next section, this latter enhancement of \hi\ deficiency might be
related to one of the classical background clouds of the Virgo cluster
region.

Previous studies by \citet*{FOY93}, \citeauthor{YFO97}, and
\citeauthor{FTS98}, among others, have shown that the Virgo spiral
distribution is strongly elongated along the LOS. The impressions
obtained above from the distribution of \hi\ deficiency, although
crude, provide further evidence for the large depth in LOS distance of
the Virgo spirals, which we now see that is also reflected in the
gaseous deficiency. Hence, in contrast to what it is commonly assumed,
not all the \hi-poor objects in the Virgo region reside in the
neighborhood of the cluster core.

\subsection{Correlation with Recessional Velocities}

The most frequently discussed gas removal mechanism that depends on the
ICM density is ram-pressure sweeping \citep{GG72,QMB00,Vol01}. For a
given galaxy, the stripping efficiency relies both on the density of
the hot intracluster gas and on the square of the relative velocity of
the galaxy with respect to the latter (actually, it is only the
component in the direction normal to the disk that matters). Suggestive
indication that it is a density that correlates with gas deficiency is
provided by the radial nature of the deficiency pattern in the core of
rich clusters \citetext{see \citeauthor{Sol01}}, which, as shown above,
it is also reproduced in the center of the Virgo cluster
region. However, attempts to unearth further evidence of ram-pressure
stripping by seeking possible correlations between \hi\ depletion and
velocity relative to the cluster are severely limited, not only by
projection effects arising from the poor correlation between the
unknown space velocities of the galaxies with the one component that
can be measured \citep[see, e.g., ][]{GH85,HG86,Mag88}, but also
because in some cases the effects of the galaxy-ICM interaction are
observed only \emph{after} the closest passage of the galaxies to the
cluster center \citep{Vol01}.

The inclusion of 2D or 3D positional information increases notably the
sensitivity of the tests. In \citeauthor{Sol01}, this approach served
to demonstrate that spirals devoid of gas follow more eccentric orbits
than the gas-rich objects in the central regions of \hi-deficient
clusters. However, in the dynamically young Virgo cluster, where the
galaxy orbits are not yet fully settled, no correlation was found
between the gas deficiency and the orbital parameters of the spirals. A
similar exercise is reproduced here in Figure~\ref{arrows} for a much
more spatially extended dataset. This figure shows the radial
velocities of the galaxies in our 21-cm sample relative to the Virgo
systemic velocity ---for this parameter we choose the typical value of
$980\,$\kms\ \citep[e.g.,][]{Tee92}--- plotted as a function of their
right ascension and LOS distance for different intervals of \hi\
deficiency. As in the previous study of the central Virgo region, the
inspection of the different panels does not reveal any clear connection
between the kinematics of the galaxies and their gas contents. An
indication exists that the galaxies with $\df=2$--$3\sigma$ closest to
the cluster core have preferentially higher relative velocities. This
impression, however, is not corroborated by the objects with the
highest gas deficiencies. The only clear trait, consistently repeated
in the four panels, is the coherence in sign and in magnitude of the
relative velocities of the galaxies located at large Virgocentric
distances, suggesting collective motions that still retain a memory of
the Hubble expansion: galaxies behind the Virgo core ($d\gtrsim
30$~Mpc) tend to move far away at high speed, while almost all galaxies
in front of this region ($d\lesssim 15$~Mpc) exhibit substantial radial
movements toward us. The feasibility of a scenario in which the
galaxies at large Virgocentric distances are deficient from having
traversed the cluster core earlier is explored in \citet{San02}.

The inclusion of spatial coordinates perpendicular to the LOS in the
present graphical analysis serves to emphasize additionally the marked
east-west asymmetry in the depth of the galaxy distribution. In the
western half of the Virgo cluster region, most galaxies have LOS
distances spread throughout the range from 10 to 50 Mpc, whereas, in
the eastern half, few objects are seen at distances larger than 25
Mpc. Interestingly enough, \citet{WB00} also detected a tendency for
the brightest elliptical galaxies located in the western region of the
Virgo cluster to be more distant than those on the eastern side. In
contrast, as evident in Figure~\ref{virgohiam}, the X-ray emission of the
cluster core is more extended towards the eastern side.

\section{The Three-Dimensional Structure of the Virgo Region}
\label{structure3d}

\begin{figure}
\epsscale{0.975} \plotone{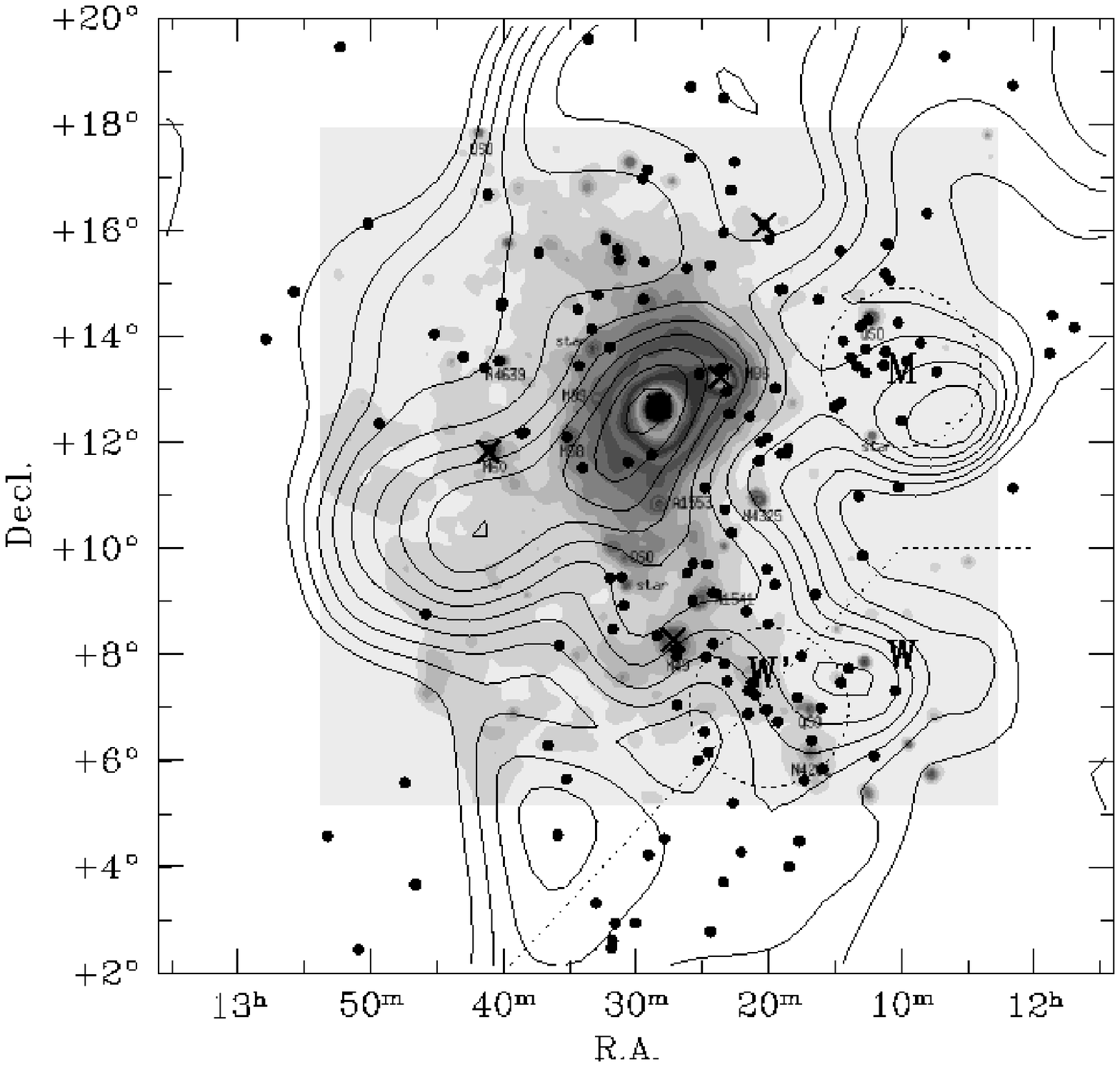} \figcaption{Distribution in
celestial coordinates of the 161 members of the 21-cm sample (five
galaxies with extreme values of declination have been omitted for a
better overall impression). The contour map of the
\emph{distance-independent} \hi\ deficiency parameter (see text) is
reproduced from \citeauthor{Sol01}, and has been generated using a
straightforward extension of the adaptive kernel technique described in
\citet{Sil86}. Note that the contours are generated from all the 287
spirals with a neutral content measure listed in the AGC within the
area depicted by the figure. A grey-scaled version of the X-ray image
of the central cluster region in the ROSAT all-sky-survey in the hard
(0.4--2.4~keV) energy band is overlaid also on the figure. The
background W, W$^\prime$, and M subgroups from \citet{BPT93} are
delineated by dotted lines. The sky positions of five dominant galaxies
are marked by crosses (top to bottom: M100, M86, M87, M60, and
M49). The projected location of M87 coincides with the peak of the \hi\
deficiency and of the X-ray emission. \label{virgohiam}}
\end{figure}

Some progress toward a precise determination of the complex structure
of the Virgo cluster region is now beginning to emerge from distance
measurement methods capable of determining individual galaxy distances
to a precision comparable to the inter-group separations. Recent
studies relying on TF, surface brightness fluctuations, or fundamental
plane distance measurement techniques \citetext{e.g.,
\citeauthor{YFO97,Gav99}; \citealt{NT00,Fou01}} have produced quite an
elaborate set of substructures and opened a debate on the original
group membership assignments of numerous galaxies. In essence, however,
they have confirmed the robustness of the original subdivision inferred
from imaging and recessional velocity data \citetext{e.g.,
\citealt{dVau61}; \citeauthor{VCC}; \citealt{BPT93}} that splits the
Virgo~I cluster region essentially in two major central subclusters and
three peripheral groups (cf.\ Fig.~\ref{virgohiam}). The largest galaxy
concentration dominates the northern part of the Virgo region and
coincides with the brightest giant elliptical, M87, which also appears
to be the center of the X-ray emission \citep{Boh94}. This main
subunit, which will be referred to here as the M87 subcluster, is
supposed to trace the cluster core, which might not be virialized
\citetext{\citealt{BPT93,Boh94}; \citealt*{SBB99}}. Another giant
elliptical, M49, marks the center of the other major Virgo galaxy
concentration, hereafter the M49 subcluster, located southwards from
the M87 subcluster. The M49 subcluster appears to be connected towards
the southwest with the W$^\prime$ and W background clouds
\citep{dVau61}, forming a continuous chain that extends up to roughly
twice the distance of the M87 subcluster (interestingly enough, a
tenuous bridge of X-ray luminous gas can be seen in
Fig.~\ref{virgohiam} connecting the M49 subcluster with the
W$^\prime$/W cloud region). Finally, in the northwest and at about the
distance of the W cloud, there is another well-defined background cloud
named M \citep*{FSF84}.

\subsection{Spherical Coordinates}\label{spherical}

Let us see now whether the main substructures of the Virgo region that
we have just enumerated bear any relationship with any of the maxima
observed in the radial pattern of the \hi\ deficiency. For this, we use
the tomographic presentation of the sky distribution of the members of
the 21-cm sample shown in Figure~\ref{tomoplot} which partitions the
galaxies into 9 segments of TF distance. 

It is seen clearly that the center of gravity of the \hi\ deficiency
distribution moves from north to south as the distance increases,
consistently following the structure of the Virgo cluster described
above. The major concentration of \hi-deficient spirals is seen in the
distance range of 15--20 Mpc encircling the position of M87. Numerous
gas-deficient objects are detected also in the panels corresponding to
the distance ranges of 10--15 Mpc and 25--30 Mpc. In the latter, these
galaxies are essentially concentrated between the southern edge of the
M49 subcluster and the W$^\prime$/W cloud region, while in the former
they tend to be located to the north of M87. Some of the gas poor
galaxies in the near distance slice could be former companions of M86
ejected at high speeds to relatively high clustercentric distances
because of the falling of this subclump into the cluster
\citetext{\citeauthor{Sol01}; \citealt{Vol01}}. The intermediate range
of $20<d<25$ Mpc is composed mainly of galaxies with moderate neutral
gas deficiencies spreaded more or less uniformly over all the
sky. Although the uncertainties in the distance estimates do not permit
a neat separation of the different Virgo substructures, it is
interesting to note that the majority of the objects in the
\hi-deficient galaxy clustering seen at 25--30 Mpc also have systemic
velocities not dissimilar from those of the M87 subcluster (see
Tables~\ref{distances} and \ref{parameters}), in agreement with the
original definition of the W$^\prime$ cloud given in VCC (note that the
W cloud is underrepresented in TF datasets). On the other hand, the
marginal indications of a galaxy enhancement in the NW of the
30--35-Mpc-distance slice might correspond to the M cloud, given that
the candidate galaxies exhibit systemic velocities around 2000
\kms. Beyond 35 Mpc, galaxies become progressively scarce, although
with an apparent tendency to reside in the peripheral W and M cloud
regions. This picture is consistent with the claims that the W and M
background clouds of Virgo are twice as far away as its central
subunits, with the W$^\prime$ cloud being somewhat closer
\citetext{e.g., \citealt{BTS87}; \citeauthor{YFO97};
\citeauthor{Gav99}}.

A final glance at Figure~\ref{tomoplot} also shows that the
gas-deficient enhancement noted in \citeauthor{Sol01} around the region
of the M cloud (see Fig.~\ref{virgohiam}) is indeed the result of the
chance superposition along the LOS of several spirals with substantial
gas deficiency, but located at very different LOS distances and without
any physical connection.

\subsection{Cartesian Coordinates}\label{xyz}

A complementary characterization of the spatial structure of the Virgo
region can be inferred from the projected distributions of the spiral
galaxies into the three main planes of the cartesian three-dimensional
space visualized in Figure~\ref{cartesian}. In the plots, the xy-plane
is taken parallel to the equatorial plane ($\rm Decl.=0\degr$), with
the x- and y-axis pointing to $\rm R.A.=12$ and 18 hr, respectively,
and the z-axis pointing to the north. In this coordinate system, the
yz-plane is nearly perpendicular to the LOS to M87, i.e., it is roughly
a tangent plane to the celestial sphere. The spherical coordinates have
been transformed into linear coordinates at the distance of the
individual galaxies listed in Table~\ref{parameters}. As in previous
figures, the symbol sizes are directly proportional to the \hi\
deficiency of the galaxies and the large cross marks the position of
M87.

This figure allows one to appreciate the true aspect of the spiral and
\hi\ distributions in the Virgo region. One remarkable feature of the
galaxy distribution is that at large radial distances ($\gtrsim 25$--30
Mpc) it appears to split into two branches in the vertical direction of
the xz-plane, which is roughly perpendicular to the plane of the Local
Supercluster. The fact that the z-axis is nearly perpendicular to the
LOS suggests that this galaxy arrangement is not an artifact produced
by the uncertainty in the radial distances affecting for the most part
the x direction. Notice also that the upper branch, which goes through
M87 and embraces the deficient objects having the most extremal radial
distances, contains nearly all the galaxies with strong \hi\
depletions, i.e., those with $\df>2\sigma$, or equivalently, with at
least a factor three reduction in the \hi\ mass. This upper filament of
the spiral distribution is pretty well aligned with the chain of bright
elliptical galaxies that defines the principal axis of Virgo
\citep{Arp68,WB00}.

To complete our description of the three-dimensional distribution of
the gaseous deficiency, we show in Figure~\ref{isosurfaces} the
surfaces of iso-\hi\ deficiency corresponding to 1, 2, and 3 standard
deviations from normalcy. These surfaces have been generated by a
straightforward extension of the adaptive kernel method described in
\citet{Sil86} for smoothing galaxy number density distributions, also
used in the determination of the \hi-deficiency contours depicted in
Figure~\ref{virgohiam}. We observe in the panel corresponding to the
$1\sigma$-level surface the presence of two disconnected condensations
of \hi\ deficiency: the elongated one in the front encompasses most of
the \hi-deficient galaxies associated with the main body of the cluster
and the gas-poor objects nearest to us, while the most distant surface
is related to the group of strongly gas-deficient galaxies that we have
tentatively identified in the background of the Virgo cluster (compare
this panel with Figs.~\ref{hivsd} and \ref{tomoplot}). The remaining
two panels illustrate the already commented fact that the highest gas
deficiencies in the Virgo region are found in the front of the
cluster. As explained in \S~\ref{bias}, little bias seems to be caused
by \hi\ deficiency in the TF distances, even for extreme gas
depletions, so the abundance of gas-poor objects at short radial
distances from our position appears to be a real effect.

\section{Summary and Remarks}\label{conclusions}

In this paper, we have examined in more detail the suggestion presented
in \citeauthor{Sol01} that, in addition to the main galaxy
concentration around M87, some of the well-known peripheral Virgo
groups also contain strongly gas-deficient spirals. The overall
distribution of \hi\ deficiency in the Virgo region has been compared
with the three-dimensional galaxy distribution. The following
conclusions have been reached:

(1) We confirm that the distribution of the spirals in the Virgo I
cluster region is very elongated along the LOS; the galaxies associated
with this region have LOS distances raging from less than 10 to more
than 50 Mpc. The projected sky distribution of the Virgo spirals,
however, looks (lumpy but) relatively compact, with a typical extent of
only about 10 Mpc. The overall width-to-depth ratio is approximately
1\,:\,4, although with a strong east-west variability. The most distant
objects concentrate in the western quadrant, while in the eastern half
few spirals are seen at LOS distances larger than 25 Mpc. The Virgo
filamentary structure appears to split into two branches around the
W$^\prime$ cloud region.

(2) The distribution of spiral galaxies with significant \hi\
deficiency is also characterized by great depth along the LOS. The
highly gas-deficient spirals tend to concentrate along the upper branch
of the spiral galaxy distribution, which is roughly aligned with the
principal axis of the Virgo cluster.

(3) Within 4 Mpc of M87, the measured \hi\ deficiency is essentially a
monotonically decreasing function of the distance from that galaxy, in
agreement with the behavior observed in other \hi-deficient
clusters. Moreover, in the Virgo region, significant \hi\ deficiency
enhancements are also identified at large distances from the Virgo
core, well beyond the typical distance where the hot X-ray emitting ICM
is concentrated. Tests of whether locally-high peripheral gas
deficiencies are a rather common feature in cluster regions must await
the equally-careful tracing of the \hi\ deficiency, incorporating
quality 3D distance measures around other clusters.

(4) While the principal peak in the distribution of \hi\ deficiency
arises from numerous gas-poor galaxies coincident with the core and
with LOS distances ranging from \sm16 to 22 Mpc, other important
enhancements of the gas deficiency are associated with several nearby
galaxies ($d\lesssim 15$ Mpc) moving away from the cluster with large
relative velocities, and with what appears to be a compact background
group of galaxies between \sm25--30 Mpc, most with roughly the same
systemic velocities as the cluster mean, which matches the original
definition of the W$^\prime$ cloud. In addition, we have demonstrated
that the localized enhancement in $\df$ observed in the Virgo sky map
around the M cloud position actually arises from several galaxies at
very different distances aligned along the LOS and without any physical
connection. In agreement with results presented by \cite*{Dal01},
nothing in our analysis suggests that TF distance measurements are
unreliable in objects with severe gas depletion.

\begin{figure}
\epsscale{1.0} \plotone{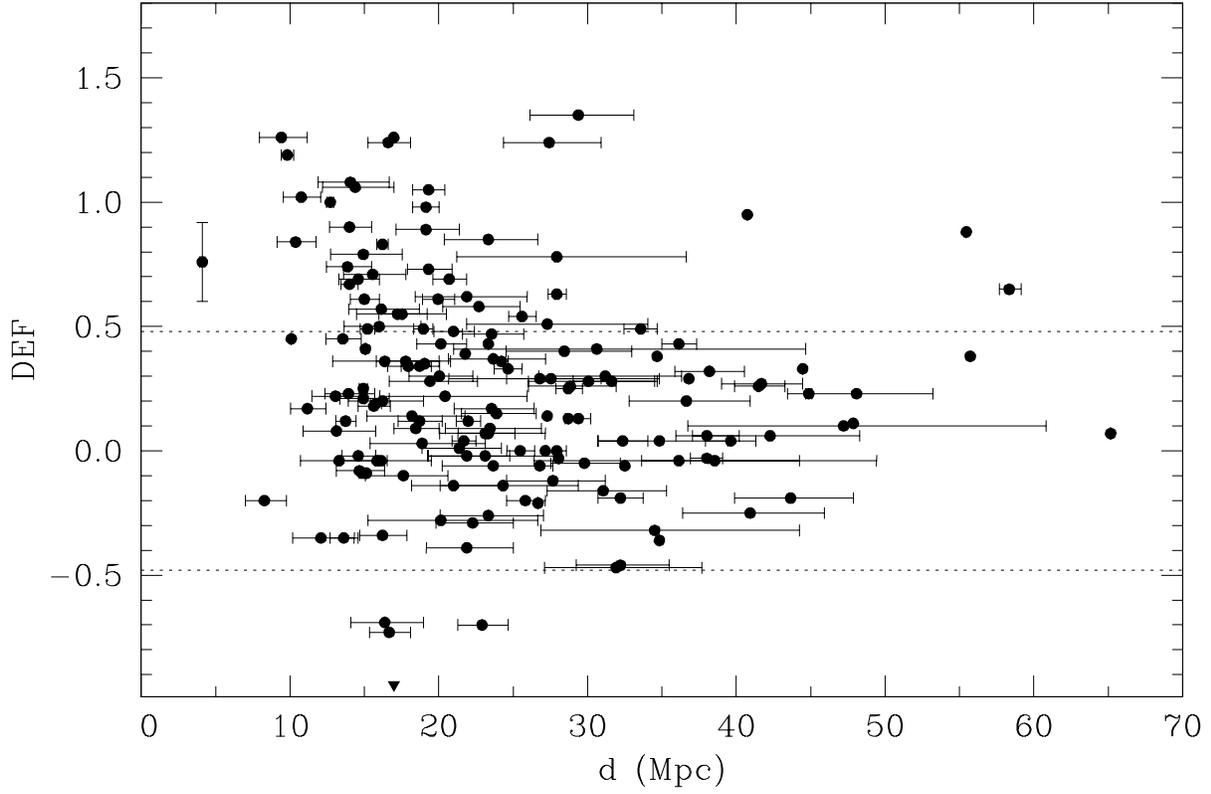} \figcaption{Individual
values of $\df$ for the 161 members of the 21-cm sample as a function
of the LOS distance. Dotted lines show 2 times the standard deviation
shown by the values of this parameter in field galaxies. Horizontal
error bars represent the $1\sigma$ uncertainties of the distances
quoted in the literature with respect to the calculated mean
values. The filled triangle marks the distance to M87 quoted in
LEDA. The vertical error bar in the point closest to us shows an
estimate of the typical uncertainty of the individual values of $\df$
expected from random errors in the determination of the observables
$a_\mathrm{opt}^2$, \fdef, and $T$, that enter in the calculation of
this parameter. \label{hivsd}}
\end{figure}

\begin{figure}
\epsscale{1.0} \plotone{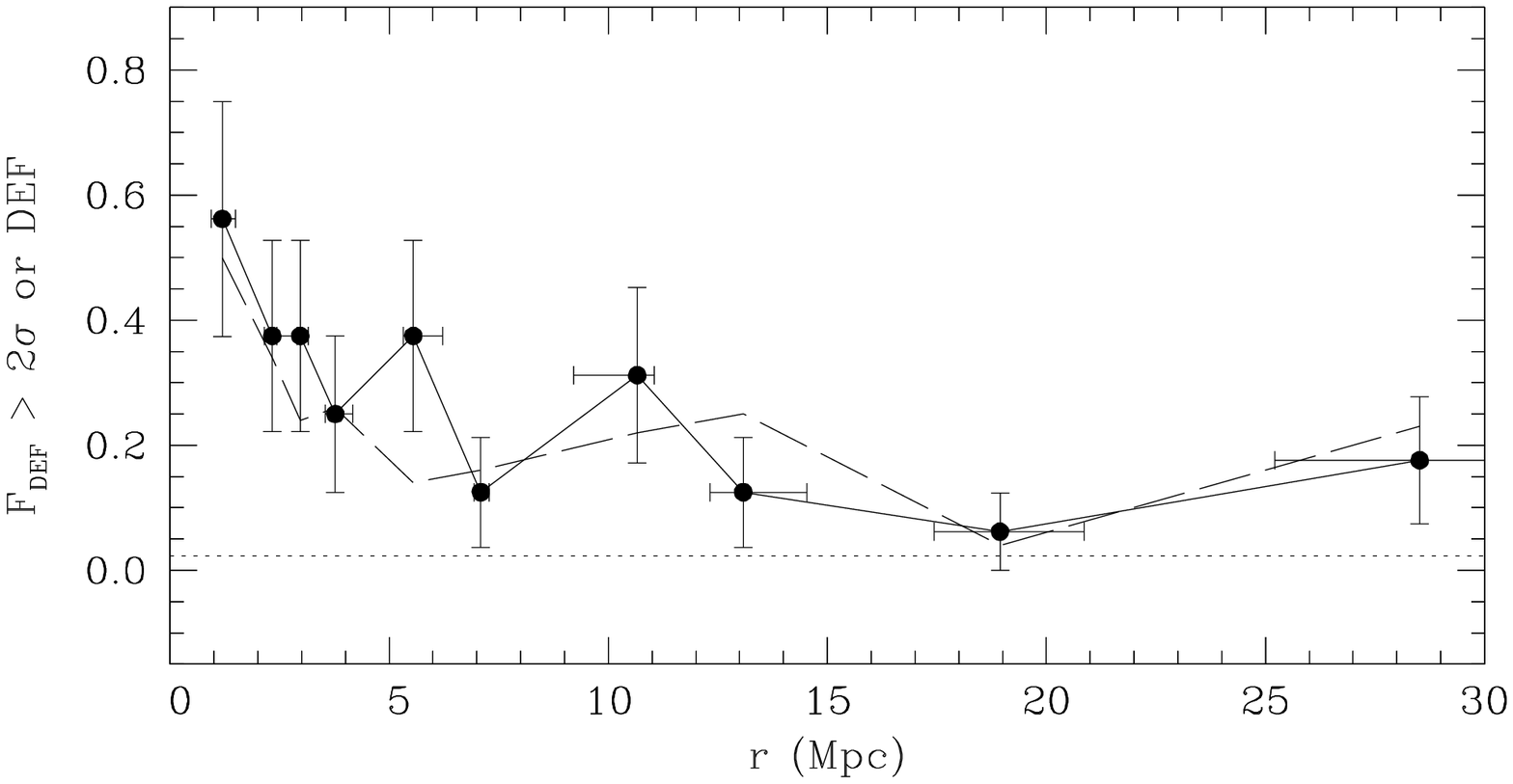} \figcaption[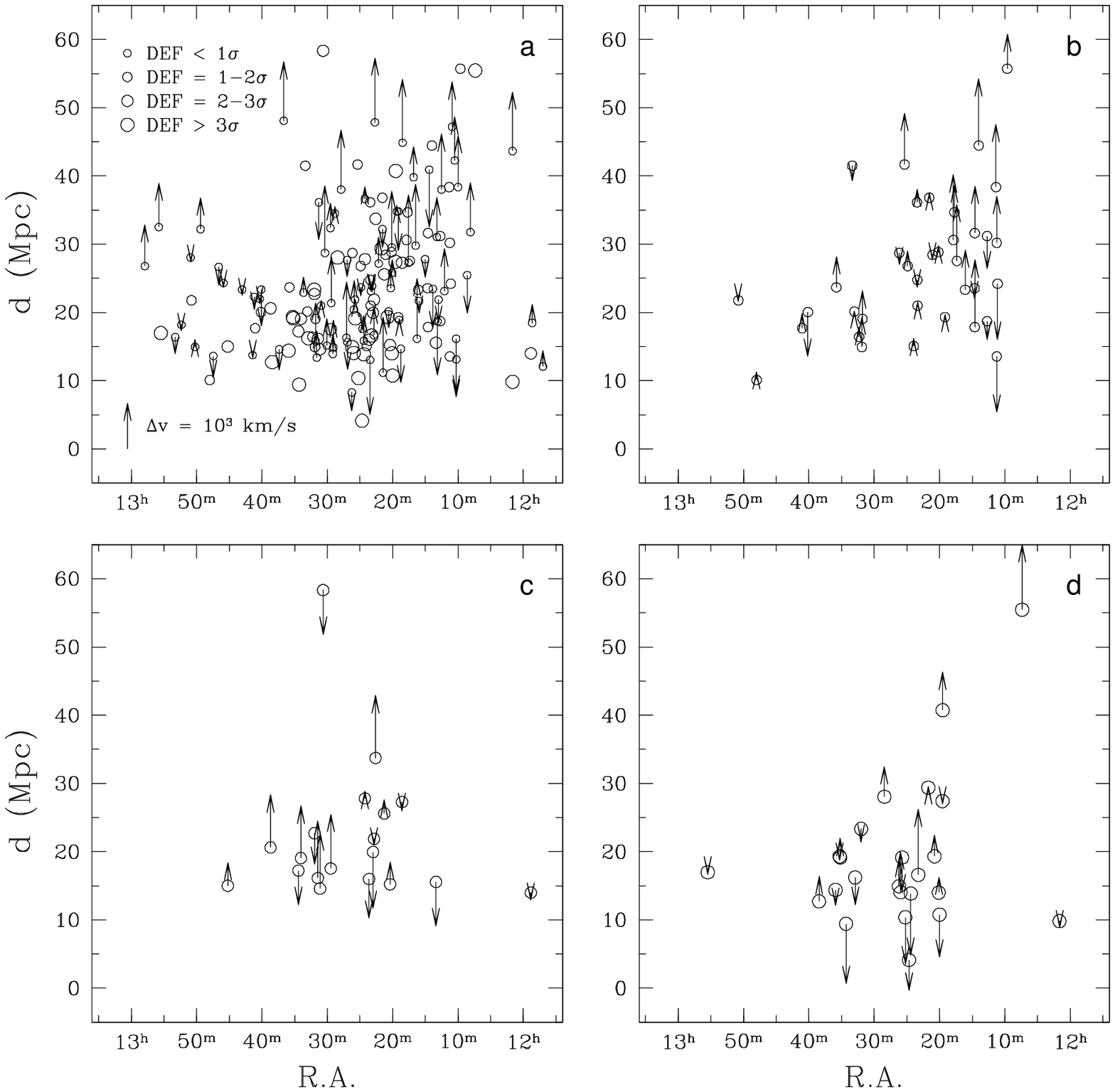]{Data points
illustrate the run of the fraction of spirals with $\df>2\sigma$ with
three-dimensional radial distance from the center of the Virgo cluster.
Vertical error bars correspond to $1\sigma$ confidence Poisson
intervals. The abscissas show medians and interquartile ranges of the
bins in distance determined from 16 galaxies, with the remainder one
added to the last bin. The horizontal dotted line is the expectation
value of \fdf\ for field spirals if $\df$ follows a gaussian
distribution. The long-dashed curve illustrates the radial run of the
medians of the binned number distributions in the measured
$\df$.\label{hivsd3}}
\end{figure}

\begin{figure}
\epsscale{1.0} \plotone{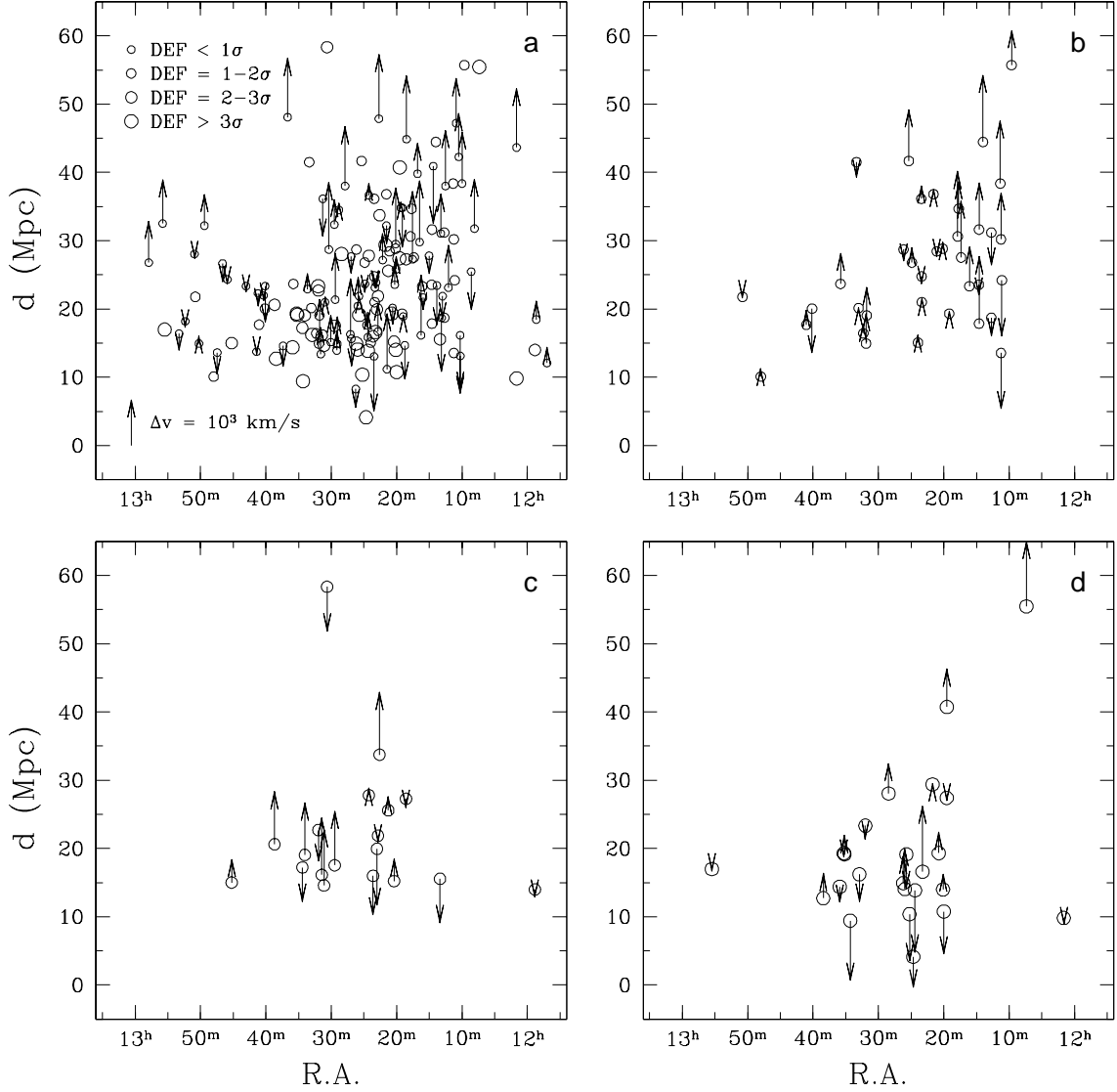} \figcaption{The distribution
of the observed relative recessional velocities for different
intervals of \hi\ deficiency. The horizontal axis shows the position of
the galaxy in Right Ascension while the vertical axis shows its LOS
distance in Mpc. The size and orientation of the arrows indicates the
radial velocity relative to the systemic cluster recessional velocity:
$\Delta v=\;$\vel$\;-\;980$ \kms. Panel (a) depicts all data points but
only the velocities for non-\hi-deficient galaxies ($\df<1\sigma$). In
the rest of panels, the \hi-deficient galaxies are separated according
to specific ranges of deficiency: (b) $\df=1$--$2\sigma$, (c)
$\df=2$--$3\sigma$, and (d) $\df>3\sigma$. Two galaxies with $d>65$ Mpc
have been excluded from the Figure.\label{arrows}}
\end{figure}

\begin{figure}
\epsscale{1.0} \plotone{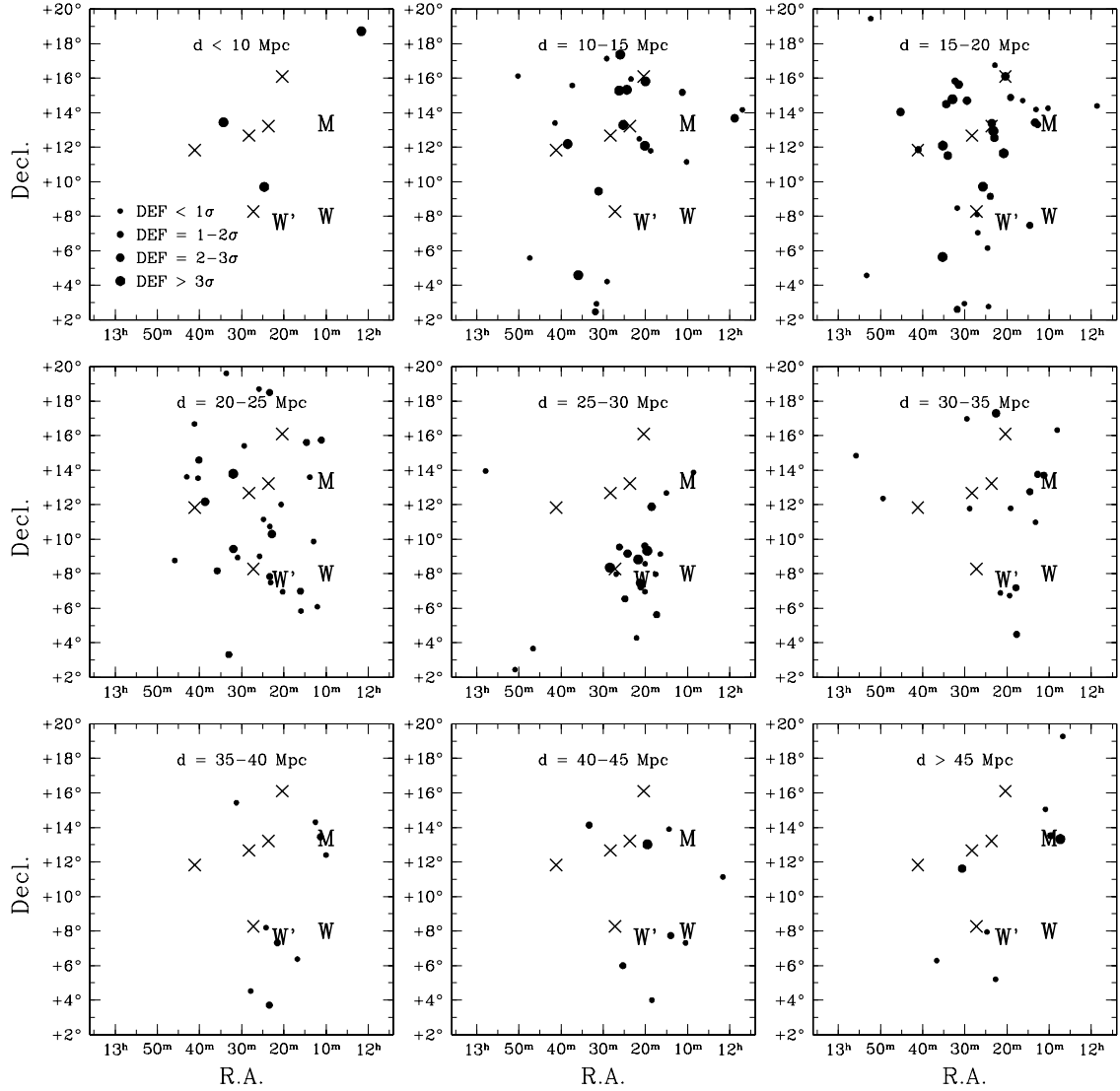} \figcaption{Sky distribution
of the Virgo spirals for specific ranges of the LOS distance. All
panels encompass a range of LOS distances of 5 Mpc, except for the
first and last ones which encompass, respectively, all the objects with
$d < 10$ Mpc and with $d > 45$ Mpc. The size of the symbols correlates
with the \hi\ deficiency of the galaxies (see the correspondences in
the first panel) measured in units of the mean standard deviation for
field objects ($=0.24$). Crosses and uppercase letters have the same
meaning as in Fig.~\ref{virgohiam}.\label{tomoplot}}
\end{figure}

\begin{figure}
\epsscale{0.425} \plotone{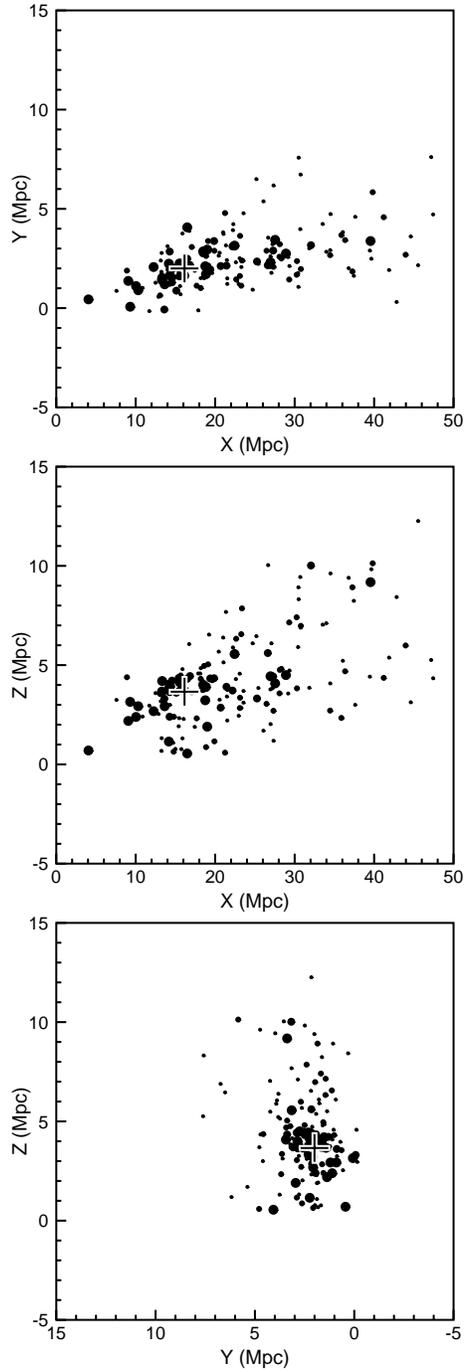} \figcaption{Distribution
of the Virgo spirals in the three main planes of the rectangular
equatorial coordinate system. The xy-plane corresponds to $\rm
Decl.=0\degr$, the x- and y-axis point to $\rm R.A.=12$ and 18 hr,
respectively, and the z-axis points to the north. As in
Figure~\ref{tomoplot}, the symbol size indicates the relative degree of
\hi\ deficiency. The large cross in each panel marks the position of
M87.\label{cartesian}}
\end{figure}

\begin{figure}
\epsscale{0.395} \plotone{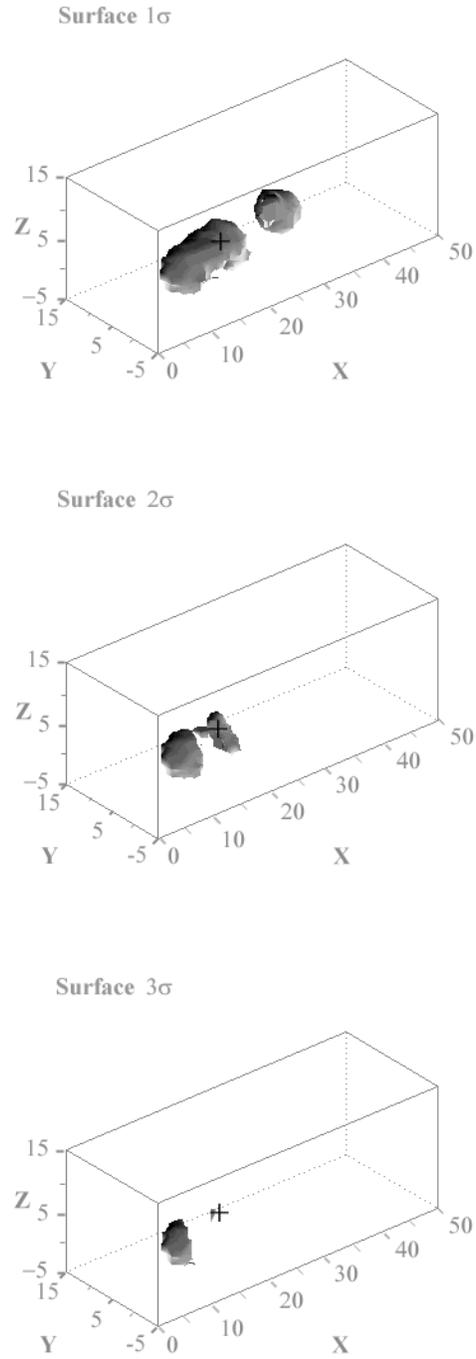} \figcaption{\emph{Top to
bottom:} Surfaces of iso-\hi\ deficiency corresponding to 1, 2, and 3
standard deviations from normalcy. All plots are in rectangular
equatorial coordinates as in Figure~\ref{cartesian} with distances
given in Mpc. The crosses mark the position of M87. Our position is at
the origin of the coordinate system. \label{isosurfaces}}
\end{figure}

Further progress in (1) and (2) needs a careful revision of TF distances
---at least until Cepheid distance measurements in Virgo galaxies
become more commonplace. Even after the elimination of systematic
differences among published Virgo catalogs, a few galaxies still
exhibit strongly inconsistent distance measurements: 16 of the 161
members of the 21-cm sample have $1\sigma$ uncertainties larger than 5
Mpc. Nevertheless, although the details may be questioned, the general
picture reporting the elongated structure of the distribution of both
the spirals and their \hi\ deficiency, as well as the clumped nature of
the latter, should be correct. We can make a simple estimate of the
\emph{typical} elongation introduced by the uncertainty in the
distances. The standard deviation of individual distances ($0.29$ mag)
is likely responsible for an increase of \sm40$\%$ in the scale of the
true LOS distance distribution by assuming an average distance of the
spiral galaxies of 20 Mpc. Given that the absolute errors in distance
increase with the values of this quantity, as it is obvious from
inspection of Figure~\ref{hivsd}, one can expect an artificial
increment of the depth by a somewhat larger factor for the most distant
galaxies which, in any event, would be clearly insufficient to account
for the very strong LOS elongation of the galaxy distribution.

On the other hand, results (3) and (4) have profound implications on
our understanding of the gas removal events and the influence of the
environment on the life of the galaxies. While the characteristics
exhibited by the \hi\ deficiency in cluster centers tend to support the
interaction between the galaxies and the hot intracluster gas as the
main cause of their gas depletion, our finding that a number of spirals
with substantial \hi\ deficiencies lie at large radial distances from
the Virgo cluster center is hard to reconcile with the proposition that
this environmental process is also the cause. At this stage, it would
be desirable to investigate whether these peripheral deficient objects
have been produced \emph{in situ} by alternative gas deficiency
mechanisms, such as galaxy-galaxy interactions, predicted to operate in
galaxy groups. In this sense, it would be of importance to perform
multi-wavelength observations of the \hi-deficient subclump detected in
the Virgo background. Model calculations show that tidal stresses in
disks generate extended tail structures in the stellar and neutral
hydrogen distributions, the latter with surface densities well above
the detection threshold of the most sensitive aperture-synthesis radio
observations. In contrast, gas depletion arising from the ram-pressure
sweeping of the interstellar medium should produce a dearth of atomic
gas in the outer portions of the disks, as well as bow shocks and dense
gaseous tails observable in X-rays.

\begin{acknowledgements}
We are especially indebted to Pascal Fouqu\'e for providing
machine-readable versions of his TF data on the Virgo cluster and for
his careful reading of an earlier version of the manuscript. Masataka
Fukugita and Martin Federspiel have also generously contributed with
their TF data to the making of our Virgo catalog. We are grateful to
all the people and institutions that have made possible the LEDA
(http://leda.univ-lyon1.fr). Sergi G\'omez assisted with one of the
figures. This work was supported by the Direcci\'on General de
Investigaci\'on Cient\'{\i}fica y T\'ecnica, under contracts PB97--0411
and AYA2000--0951. J.M.S.\ would like also to thank the Departament
d'Astronomia i Meteorologia at the Universitat de Barcelona where part 
of this work was prepared. T.S.\ acknowledges support from a fellowship
of the Ministerio de Educaci\'on, Cultura y Deporte of Spain. This work
has been supported of US NSF grants AST96--17069 to R.G.\ and
AST99--00695 to M.P.H.\ and R.G.
\end{acknowledgements}


\clearpage
\onecolumn

\clearpage 




\begin{deluxetable}{llrr}
\tablewidth{0pt}
\tablenum{1}
\tablecaption{TF Datasets Contributing to the Present Sample \label{datasets}}
\tablecolumns{4}
\tablehead{ 
\multicolumn{1}{c}{} & \multicolumn{1}{c}{} & \multicolumn{2}{c}{Number of galaxies}\\
\multicolumn{1}{c}{Source} & \multicolumn{1}{c}{Acronym} & \multicolumn{1}{c}{\phn Total} &
\multicolumn{1}{c}{Selected}
}
\startdata
\citealt*{YFO97} & \phn \citeauthor{YFO97} & 246\phn & 165\phn\phn\phn\\
\citealt*{MAH80} & \phn \citeauthor{MAH80} & 23\phn & 21\phn\phn\phn\\
\citealt*{PT88} & \phn \citeauthor{PT88} & 34\phn & 34\phn\phn\phn\\
\citealt*{KCT88} & \phn \citeauthor{KCT88} & 128\phn & 84\phn\phn\phn\\
\citealt*{Fou90} & \phn \citeauthor{Fou90} & 178\phn & 145\phn\phn\phn\\
\citealt*{FTS98} & \phn \citeauthor{FTS98} & 132\phn & 129\phn\phn\phn\\
\citealt*{Gav99} & \phn \citeauthor{Gav99} & 75\phn & 72\phn\phn\phn\\
\citealt*{Ekh00} & \phn \citeauthor{Ekh00} & 96\phn & 52\phn\phn\phn\\
\enddata
\end{deluxetable}

\clearpage


\clearpage



\newcommand{\nod}{\nodata}
\newcommand{\noc}{\multicolumn{1}{c}{\nod}}
\begin{deluxetable}{lllccrc@{\extracolsep{-3pt}}c@{\extracolsep{-3pt}}c@
{\extracolsep{-3pt}}c@{\extracolsep{-3pt}}c@{\extracolsep{-3pt}}c@{\extracolsep{-3pt}}c@
{\extracolsep{-3pt}}c@{\extracolsep{-3pt}}c@{\extracolsep{-3pt}}cl}
\rotate
\tablewidth{715.5986pt}
\tablenum{2}
\tablecaption{Distance Moduli for Individual Galaxies \label{distances}}
\tablecolumns{17}
\tablehead{ 
& \multicolumn{1}{c}{NGC/} & \multicolumn{1}{c}{UGC/} & \multicolumn{1}{c}{R.A.} & \multicolumn{1}{c}{Decl.} & 
& & & & & & & & & & & \multicolumn{1}{c}{Distance}\\
\multicolumn{1}{c}{VCC} & \multicolumn{1}{c}{IC} & \multicolumn{1}{c}{CGCG} &
\multicolumn{1}{c}{(1950)} & \multicolumn{1}{c}{(1950)} & 
\multicolumn{1}{c}{T} & \multicolumn{1}{c}{Mem.} &
\multicolumn{1}{c}{\citeauthor{YFO97}} & \multicolumn{1}{c}{\citeauthor{MAH80}} & \multicolumn{1}{c}{\citeauthor{PT88}} &
\multicolumn{1}{c}{\citeauthor{KCT88}} & \multicolumn{1}{c}{\citeauthor{Fou90}} & \multicolumn{1}{c}{\citeauthor{FTS98}} &
\multicolumn{1}{c}{\citeauthor{Gav99}} & \multicolumn{1}{c}{\citeauthor{Ekh00}} & \multicolumn{1}{c}{Ceph.} &
\multicolumn{1}{c}{modulus}\\
\multicolumn{1}{c}{(1)} & \multicolumn{1}{c}{(2)} & \multicolumn{1}{c}{(3)} &
\multicolumn{1}{c}{(4)} & \multicolumn{1}{c}{(5)} &
\multicolumn{1}{c}{(6)} & \multicolumn{1}{c}{(7)} &
\multicolumn{1}{c}{(8)} & \multicolumn{1}{c}{(9)} & \multicolumn{1}{c}{(10)} &
\multicolumn{1}{c}{(11)} & \multicolumn{1}{c}{(12)} & \multicolumn{1}{c}{(13)} &
\multicolumn{1}{c}{(14)} & \multicolumn{1}{c}{(15)} & \multicolumn{1}{c}{(16)} &
\multicolumn{1}{c}{(17)}
}
\startdata
\noc & \noc & CG69-10& 11\phn 57\phn 00.8&$+$14\phn 09\phn 51 & 7 & \nod & 30.82 & \nod & \nod & \nod & 29.83 & \nod & \nod & \nod & \nod & $30.41\pm 0.37$\\
\noc & I0755 & U7001 & 11\phn 58\phn 37.3&$+$14\phn 23\phn 13 & 3 & \nod & 31.63 & \nod & \nod & \nod & 31.03 & 31.20 & \nod & \nod & \nod & $31.33\pm 0.18$\\
\noc & N4037 & U7002 & 11\phn 58\phn 50.6&$+$13\phn 40\phn 41 & 3 & \nod & 30.68 & \nod & \nod & \nod & 30.61 & \nod & \nod & \nod & \nod & $30.73\pm 0.09$\\
\noc & N4064 & U7054 & 12\phn 01\phn 37.3&$+$18\phn 43\phn 16 & 1 & \nod & 29.94 & \nod & \nod & 30.18 & \nod & 29.97 & \nod & \nod & \nod & $29.96\pm 0.09$\\
\noc & N4067 & U7048 & 12\phn 01\phn 38.2&$+$11\phn 08\phn 13 & 3 & \nod & 33.39 & \nod & \nod & \nod & 32.70 & 33.29 & \nod & \nod & \nod & $33.20\pm 0.20$\\
\noc & \noc & U7133 & 12\phn 06\phn 46.7&$+$19\phn 16\phn 31 & 7 & \nod & \nod & \nod & \nod & \nod & \nod & 33.99 & \nod & \nod & \nod & $34.07        $\\
V0015 & I3021 & U7149 & 12\phn 07\phn 21.6&$+$13\phn 19\phn 42 & 9 & M & 33.79 & \nod & \nod & \nod & \nod & \nod & \nod & \nod & \nod & $33.72        $\\
V0025 & N4152 & U7169 & 12\phn 08\phn 03.8&$+$16\phn 18\phn 45 & 5 & \nod & 32.27 & \nod & \nod & \nod & 32.00 & \nod & 32.85 & 33.05 & \nod & $32.52\pm 0.36$\\
V0034 & I3033 & U7181 & 12\phn 08\phn 37.2&$+$13\phn 51\phn 54 & 7 & A & 32.12 & \nod & \nod & 31.99 & 31.88 & 32.04 & \nod & \nod & \nod & $32.03\pm 0.08$\\
V0047 & N4165 & U7201 & 12\phn 09\phn 39.0&$+$13\phn 31\phn 30 & 2 & M & 33.80 & \nod & \nod & \nod & \nod & \nod & \nod & \nod & \nod & $33.73        $\\
V0058 & I0769 & U7209 & 12\phn 09\phn 59.4&$+$12\phn 24\phn 00 & 4 & M & 32.39 & \nod & 32.68 & \nod & \nod & 32.99 & 32.92 & 33.52 & \nod & $32.93\pm 0.30$\\
V0066 & N4178 & U7215 & 12\phn 10\phn 13.8&$+$11\phn 08\phn 48 & 7 & \nod & 30.51 & 31.02 & 30.52 & 29.44 & 30.22 & 30.78 & 30.86 & 30.59 & \nod & $30.59\pm 0.40$\\
V0067 & I3044 & U7216 & 12\phn 10\phn 15.6&$+$14\phn 15\phn 18 & 6 & A & 31.57 & \nod & \nod & 30.69 & 30.94 & 30.94 & \nod & \nod & \nod & $31.05\pm 0.34$\\
V0073 & N4180 & U7219 & 12\phn 10\phn 28.9&$+$07\phn 19\phn 01 & 2 & W & 33.33 & \nod & \nod & \nod & 32.50 & 33.53 & 32.72 & 33.57 & \nod & $33.13\pm 0.29$\\
V0081 & \noc & U7223 & 12\phn 10\phn 53.0&$+$15\phn 03\phn 00 & 7 & A & 32.87 & \nod & \nod & \nod & 33.70 & \nod & \nod & \nod & \nod & $33.37\pm 0.55$\\
V0087 & \noc & \noc & 12\phn 11\phn 07.0&$+$15\phn 43\phn 54 & 9 & A & 31.92 & \nod & \nod & 32.02 & 31.75 & \nod & \nod & \nod & \nod & $31.92\pm 0.04$\\
V0089 & N4189 & U7235 & 12\phn 11\phn 14.4&$+$13\phn 42\phn 12 & 6 & M & 32.64 & \nod & 31.63 & \nod & \nod & 32.64 & \nod & 33.01 & \nod & $32.39\pm 0.31$\\
V0092 & N4192 & U7231 & 12\phn 11\phn 15.4&$+$15\phn 10\phn 23 & 3 & A & 30.81 & 30.67 & 30.65 & 30.55 & 30.59 & 30.98 & 30.36 & 31.17 & \nod & $30.66\pm 0.19$\\
V0097 & N4193 & U7234 & 12\phn 11\phn 21.0&$+$13\phn 27\phn 00 & 4 & M & 32.78 & \nod & \nod & \nod & \nod & 32.98 & 32.64 & 33.54 & \nod & $32.91\pm 0.13$\\
V0105 & \noc & U7239 & 12\phn 11\phn 36.0&$+$08\phn 03\phn 00 & 10& \nod & 32.08 & \nod & \nod & \nod & 32.61 & \nod & \nod & \nod & \nod & $32.43\pm 0.40$\\
V0119 & \noc & U7249 & 12\phn 12\phn 05.4&$+$13\phn 05\phn 24 & 10& A & 31.86 & \nod & 31.78 & 31.89 & 31.62 & 31.78 & \nod & \nod & \nod & $31.85\pm 0.10$\\
V0120 & N4197 & U7247 & 12\phn 12\phn 04.9&$+$06\phn 05\phn 01 & 6 & \nod & 31.82 & \nod & \nod & 31.89 & 31.68 & 31.56 & 31.91 & \nod & \nod & $31.82\pm 0.18$\\
V0126 & I3059 & U7254 & 12\phn 12\phn 22.8&$+$13\phn 44\phn 12 & 10& A & 31.87 & \nod & \nod & \nod & 31.73 & 31.92 & \nod & \nod & \nod & $31.89\pm 0.05$\\
V0131 & I3061 & U7255 & 12\phn 12\phn 31.8&$+$14\phn 18\phn 24 & 5 & M & 33.11 & \nod & \nod & \nod & 32.75 & 32.70 & \nod & 33.37 & \nod & $32.90\pm 0.12$\\
V0132 & \noc & \noc & 12\phn 12\phn 32.3&$+$13\phn 18\phn 41 & 8 & A & 30.67 & \nod & \nod & \nod & \nod & 29.77 & \nod & \nod & \nod & $30.16\pm 0.47$\\
V0143 & I3066 & U7262 & 12\phn 12\phn 43.2&$+$13\phn 45\phn 06 & 4 & A & 32.89 & \nod & \nod & \nod & 32.33 & 32.04 & \nod & \nod & \nod & $32.47\pm 0.33$\\
V0145 & N4206 & U7260 & 12\phn 12\phn 44.4&$+$13\phn 18\phn 12 & 4 & A & 31.48 & 31.36 & 31.01 & 31.47 & 31.20 & 31.15 & 31.29 & \nod & \nod & $31.36\pm 0.15$\\
V0152 & N4207 & U7268 & 12\phn 12\phn 58.2&$+$09\phn 51\phn 48 & 6 & \nod & 31.87 & \nod & \nod & 32.00 & 31.65 & 31.75 & 31.08 & \nod & \nod & $31.70\pm 0.27$\\
V0157 & N4212 & U7275 & 12\phn 13\phn 06.6&$+$14\phn 10\phn 48 & 5 & A & 31.36 & \nod & 31.45 & 31.54 & 31.23 & 31.23 & 31.04 & 31.98 & \nod & $31.36\pm 0.17$\\
V0162 & I3074 & U7279 & 12\phn 13\phn 13.4&$+$10\phn 58\phn 36 & 8 & \nod & 32.51 & \nod & 32.64 & 32.49 & 32.01 & 32.21 & \nod & \nod & \nod & $32.46\pm 0.28$\\
V0167 & N4216 & U7284 & 12\phn 13\phn 21.6&$+$13\phn 25\phn 36 & 3 & A & 31.13 & 30.81 & 30.81 & 30.73 & 31.23 & 31.35 & 30.65 & 31.34 & \nod & $30.96\pm 0.29$\\
V0187 & N4222 & U7291 & 12\phn 13\phn 49.8&$+$13\phn 35\phn 12 & 6 & A & 31.80 & \nod & \nod & 31.87 & 31.70 & 31.46 & \nod & 32.91 & \nod & $31.85\pm 0.30$\\
V0199 & N4224 & U7292 & 12\phn 14\phn 00.4&$+$07\phn 44\phn 20 & 1 & W & 33.30 & \nod & \nod & \nod & \nod & \nod & \nod & \nod & \nod & $33.24        $\\
V0213 & I3094 & U7305 & 12\phn 14\phn 23.4&$+$13\phn 54\phn 12 & 5 & A & 33.37 & \nod & \nod & \nod & 32.60 & \nod & \nod & \nod & \nod & $33.06\pm 0.25$\\
V0222 & N4235 & U7310 & 12\phn 14\phn 35.7&$+$07\phn 28\phn 11 & 1 & W & 31.48 & \nod & \nod & \nod & \nod & \nod & \nod & 31.76 & \nod & $31.27\pm 0.18$\\
V0224 & I3099 & U7313 & 12\phn 14\phn 37.2&$+$12\phn 44\phn 54 & 6 & M & 32.39 & \nod & \nod & \nod & 32.55 & 32.37 & \nod & \nod & \nod & $32.50\pm 0.19$\\
V0226 & N4237 & U7315 & 12\phn 14\phn 38.2&$+$15\phn 36\phn 08 & 4 & A & 31.63 & \nod & \nod & 32.15 & 31.80 & 31.95 & 31.48 & 32.52 & \nod & $31.86\pm 0.19$\\
V0241 & I3105 & U7326 & 12\phn 15\phn 01.2&$+$12\phn 40\phn 00 & 10& A & 30.48 & \nod & \nod & 30.56 & 30.74 & 30.69 & \nod & \nod & \nod & $30.62\pm 0.20$\\
V0267 & I3115 & U7333 & 12\phn 15\phn 26.4&$+$06\phn 55\phn 53 & 6 & \nod & 34.92 & \nod & \nod & \nod & 31.96 & \nod & 32.08 & \nod & \nod & $32.23\pm 0.05$\\
V0289 & N4252 & U7343 & 12\phn 15\phn 57.6&$+$05\phn 50\phn 18 & 3 & \nod & 31.81 & \nod & \nod & \nod & \nod & 31.62 & \nod & \nod & \nod & $31.68\pm 0.08$\\
V0297 & \noc & CG42-1& 12\phn 16\phn 05.4&$+$06\phn 59\phn 06 & 5 & \nod & \nod & \nod & \nod & \nod & \nod & 31.85 & \nod & \nod & \nod & $31.84        $\\
V0307 & N4254 & U7345 & 12\phn 16\phn 16.8&$+$14\phn 41\phn 42 & 5 & A & 30.56 & \nod & 31.28 & \nod & 30.54 & 31.42 & \nod & \nod & \nod & $31.04\pm 0.41$\\
V0318 & I0776 & U7352 & 12\phn 16\phn 30.0&$+$09\phn 08\phn 06 & 8 & \nod & 32.36 & \nod & 32.24 & 32.27 & 31.99 & 32.58 & \nod & \nod & \nod & $32.37\pm 0.17$\\
V0341 & N4260 & U7361 & 12\phn 16\phn 48.8&$+$06\phn 22\phn 40 & 1 & W & 33.14 & \nod & \nod & \nod & \nod & \nod & \nod & 33.42 & \nod & $32.99\pm 0.09$\\
V0343 & I3148 & \noc & 12\phn 16\phn 48.0&$+$08\phn 09\phn 00 & 8 & W & 31.93 & \nod & \nod & \nod & \nod & \nod & \nod & \nod & \nod & $31.88        $\\
V0382 & N4273 & U7380 & 12\phn 17\phn 22.3&$+$05\phn 37\phn 27 & 5 & W & 31.83 & \nod & \nod & \nod & \nod & 32.27 & 32.19 & 32.91 & \nod & $32.20\pm 0.25$\\
V0393 & N4276 & U7385 & 12\phn 17\phn 34.7&$+$07\phn 58\phn 10 & 6 & W & 32.23 & \nod & \nod & \nod & \nod & \nod & \nod & \nod & \nod & $32.18        $\\
V0404 & \noc & U7387 & 12\phn 17\phn 42.9&$+$04\phn 28\phn 47 & 7 & W & \nod & \nod & \nod & \nod & \nod & 32.67 & \nod & \nod & \nod & $32.70        $\\
V0415 & \noc & CG42-36 & 12\phn 17\phn 52.8&$+$07\phn 11\phn 12 & 5 & W & 33.31 & \nod & \nod & \nod & \nod & 31.63 & \nod & \nod & \nod & $32.43\pm 0.82$\\
V0449 & N4289 & U7403 & 12\phn 18\phn 30.0&$+$04\phn 00\phn 00 & 6 & W & \nod & \nod & \nod & \nod & \nod & 33.20 & \nod & 33.75 & \nod & $33.26\pm 0.01$\\
V0453 & \noc & \noc & 12\phn 18\phn 33.0&$+$11\phn 52\phn 24 & 6 & A & 32.74 & \nod & \nod & 32.42 & 31.31 & \nod & \nod & \nod & \nod & $32.18\pm 0.48$\\
V0460 & N4293 & U7405 & 12\phn 18\phn 41.1&$+$18\phn 39\phn 36 & 0 & A & 31.14 & \nod & \nod & 30.81 & 30.91 & \nod & 30.53 & 31.03 & \nod & $30.76\pm 0.32$\\
V0465 & N4294 & U7407 & 12\phn 18\phn 45.0&$+$11\phn 47\phn 24 & 6 & A & 30.66 & 31.15 & \nod & 30.85 & 30.55 & 30.66 & 30.86 & \nod & \nod & $30.83\pm 0.24$\\
V0483 & N4298 & U7412 & 12\phn 19\phn 00.6&$+$14\phn 53\phn 06 & 5 & A & 31.03 & \nod & \nod & 32.22 & 31.04 & 31.19 & \nod & \nod & \nod & $31.38\pm 0.44$\\
V0491 & N4299 & U7414 & 12\phn 19\phn 07.8&$+$11\phn 46\phn 48 & 8 & A & \nod & \nod & \nod & \nod & 32.49 & \nod & \nod & \nod & \nod & $32.71        $\\
V0497 & N4302 & U7418 & 12\phn 19\phn 10.2&$+$14\phn 52\phn 36 & 5 & A & \nod & \nod & \nod & 31.82 & 31.53 & 31.51 & 30.78 & 32.01 & \nod & $31.44\pm 0.33$\\
V0509 & \noc & U7423 & 12\phn 19\phn 22.3&$+$06\phn 43\phn 43 & 9 & W' & 32.67 & \nod & \nod & 32.67 & 32.27 & 33.13 & \nod & \nod & \nod & $32.71\pm 0.27$\\
V0512 & \noc & U7421 & 12\phn 19\phn 23.0&$+$12\phn 14\phn 35 & 10& A & 32.05 & \nod & \nod & 32.02 & 31.30 & \nod & \nod & \nod & \nod & $31.82\pm 0.21$\\
V0514 & \noc & U7424 & 12\phn 19\phn 25.2&$+$08\phn 57\phn 06 & 10& B & 32.00 & \nod & \nod & \nod & 30.70 & \nod & \nod & \nod & \nod & $31.43\pm 0.52$\\
V0522 & N4305 & U7432 & 12\phn 19\phn 31.2&$+$13\phn 01\phn 06 & 1 & A & \nod & \nod & \nod & 33.15 & \nod & \nod & \nod & \nod & \nod & $33.05        $\\
V0524 & N4307 & U7431 & 12\phn 19\phn 33.0&$+$09\phn 19\phn 06 & 3 & B & \nod & \nod & \nod & 32.75 & 31.91 & 32.02 & 32.05 & 32.49 & \nod & $32.19\pm 0.26$\\
V0534 & N4309 & U7435 & 12\phn 19\phn 38.9&$+$07\phn 25\phn 20 & -1& W' & 32.26 & \nod & \nod & \nod & 31.61 & \nod & \nod & \nod & \nod & $32.02\pm 0.19$\\
V0559 & N4312 & U7442 & 12\phn 19\phn 59.4&$+$15\phn 48\phn 58 & 2 & A & 30.46 & \nod & \nod & \nod & 30.03 & \nod & 29.84 & \nod & \nod & $30.16\pm 0.25$\\
V0566 & \noc & \noc & 12\phn 20\phn 05.0&$+$08\phn 34\phn 24 & 9 & W' & 32.17 & \nod & \nod & 32.25 & 31.68 & \nod & \nod & \nod & \nod & $32.06\pm 0.11$\\
V0567 & I3225 & U7441 & 12\phn 20\phn 06.5&$+$06\phn 57\phn 14 & 8 & W & 32.33 & \nod & \nod & \nod & \nod & 32.40 & \nod & \nod & \nod & $32.34\pm 0.06$\\
V0570 & N4313 & U7445 & 12\phn 20\phn 06.6&$+$12\phn 04\phn 54 & 2 & A & 30.79 & \nod & \nod & 30.72 & 30.82 & 30.91 & 30.34 & \nod & \nod & $30.73\pm 0.22$\\
V0576 & N4316 & U7447 & 12\phn 20\phn 10.2&$+$09\phn 36\phn 36 & 6 & B & 32.42 & \nod & \nod & 32.55 & 32.30 & 32.28 & 31.70 & 32.88 & \nod & $32.30\pm 0.22$\\
V0593 & I3229 & U7448 & 12\phn 20\phn 18.0&$+$06\phn 57\phn 00 & 4 & W' & 32.07 & \nod & \nod & \nod & 31.83 & 31.52 & \nod & \nod & \nod & $31.86\pm 0.25$\\
V0596 & N4321 & U7450 & 12\phn 20\phn 23.2&$+$16\phn 06\phn 00 & 4 & A & 30.45 & \nod & 31.32 & \nod & 30.41 & 30.91 & 30.82 & 30.94 & 30.91 & $30.91\pm 0.07$\\
V0613 & N4324 & U7451 & 12\phn 20\phn 32.5&$+$05\phn 31\phn 36 & -1& W & 31.81 & \nod & \nod & \nod & \nod & \nod & 31.19 & \nod & \nod & $31.53\pm 0.23$\\
V0620 & I3239 & \noc & 12\phn 20\phn 37.8&$+$12\phn 00\phn 18 & 9 & A & 32.36 & \nod & \nod & 31.49 & 30.63 & \nod & \nod & \nod & \nod & $31.52\pm 0.61$\\
V0630 & N4330 & U7456 & 12\phn 20\phn 45.0&$+$11\phn 38\phn 42 & 6 & A & 31.21 & \nod & \nod & 31.76 & 31.30 & 31.38 & 31.31 & \nod & \nod & $31.43\pm 0.17$\\
V0656 & N4343 & U7465 & 12\phn 21\phn 05.0&$+$07\phn 13\phn 58 & 3 & W' & 32.75 & \nod & \nod & \nod & 32.24 & 32.39 & 31.71 & 32.54 & \nod & $32.27\pm 0.32$\\
V0664 & I3258 & U7470 & 12\phn 21\phn 12.0&$+$12\phn 45\phn 18 & 10& A & 30.18 & \nod & \nod & \nod & 30.50 & 31.57 & \nod & \nod & \nod & $30.80\pm 0.58$\\
V0667 & I3259 & U7469 & 12\phn 21\phn 16.2&$+$07\phn 27\phn 49 & 8 & W' & 32.00 & \nod & \nod & \nod & 31.80 & 32.14 & \nod & \nod & \nod & $32.04\pm 0.08$\\
V0688 & N4353 & \noc & 12\phn 21\phn 24.0&$+$08\phn 04\phn 00 & 10& B & 32.84 & \nod & \nod & \nod & 32.05 & 32.68 & \nod & \nod & \nod & $32.59\pm 0.23$\\
V0692 & N4351 & U7476 & 12\phn 21\phn 29.4&$+$12\phn 28\phn 54 & 2 & A & 30.16 & \nod & \nod & \nod & 30.18 & 30.01 & 30.48 & \nod & \nod & $30.24\pm 0.23$\\
V0697 & I3267 & U7474 & 12\phn 21\phn 33.0&$+$07\phn 19\phn 12 & 6 & W' & 32.89 & \nod & \nod & \nod & \nod & \nod & \nod & \nod & \nod & $32.83        $\\
V0699 & I3268 & U7477 & 12\phn 21\phn 34.7&$+$06\phn 53\phn 00 & 5 & B & 32.81 & \nod & \nod & \nod & 32.11 & \nod & \nod & \nod & \nod & $32.54\pm 0.21$\\
V0713 & N4356 & U7482 & 12\phn 21\phn 42.1&$+$08\phn 48\phn 48 & 6 & B & 31.96 & \nod & \nod & 32.56 & 32.37 & 32.36 & \nod & \nod & \nod & $32.34\pm 0.26$\\
V0737 & \noc & CG42-86 & 12\phn 22\phn 06.6&$+$04\phn 16\phn 36 & 8 & W & \nod & \nod & \nod & \nod & \nod & 32.17 & \nod & \nod & \nod & $32.17        $\\
V0740 & \noc & \noc & 12\phn 22\phn 07.0&$+$08\phn 46\phn 47 & 10& B & 33.22 & \nod & \nod & 32.75 & 32.21 & \nod & \nod & \nod & \nod & $32.75\pm 0.30$\\
V0768 & I3298 & \noc & 12\phn 22\phn 36.0&$+$17\phn 17\phn 00 & 4 & \nod & 32.79 & \nod & \nod & \nod & 32.34 & 32.59 & \nod & \nod & \nod & $32.63\pm 0.07$\\
V0785 & N4378 & U7497 & 12\phn 22\phn 44.3&$+$05\phn 12\phn 13 & 1 & W & 33.46 & \nod & \nod & \nod & \nod & \nod & \nod & \nod & \nod & $33.40        $\\
V0787 & N4376 & U7498 & 12\phn 22\phn 45.3&$+$06\phn 01\phn 06 & 10& \nod & 31.49 & \nod & \nod & 31.31 & 31.13 & 31.55 & 31.89 & \nod & \nod & $31.52\pm 0.29$\\
V0792 & N4380 & U7503 & 12\phn 22\phn 49.8&$+$10\phn 17\phn 36 & 2 & B & 31.47 & 31.20 & 31.30 & 31.27 & 32.09 & 32.13 & 31.73 & 32.42 & \nod & $31.70\pm 0.37$\\
V0801 & N4383 & U7507 & 12\phn 22\phn 53.8&$+$16\phn 44\phn 48 & 1 & A & 30.90 & \nod & \nod & \nod & 30.95 & \nod & 31.19 & \nod & \nod & $31.11\pm 0.18$\\
V0809 & I3311 & U7510 & 12\phn 23\phn 00.6&$+$12\phn 32\phn 12 & 7 & A & 31.62 & \nod & \nod & 31.50 & 31.43 & 31.38 & \nod & \nod & \nod & $31.50\pm 0.12$\\
V0827 & I3322A& U7513 & 12\phn 23\phn 09.9&$+$07\phn 29\phn 36 & 6 & B & 31.94 & 31.51 & 31.74 & 32.18 & 32.12 & 31.66 & 31.58 & 32.29 & \nod & $31.89\pm 0.23$\\
V0836 & N4388 & U7520 & 12\phn 23\phn 13.8&$+$12\phn 56\phn 18 & 3 & A & 31.11 & 31.09 & 30.75 & 30.84 & 30.89 & 31.13 & 31.13 & 32.05 & \nod & $31.10\pm 0.19$\\
V0849 & N4390 & U7519 & 12\phn 23\phn 19.8&$+$10\phn 43\phn 48 & 5 & A & 31.45 & \nod & \nod & \nod & 31.21 & 32.15 & 32.09 & \nod & \nod & $31.82\pm 0.40$\\
V0851 & I3322 & U7518 & 12\phn 23\phn 21.6&$+$07\phn 50\phn 00 & 6 & B & 31.72 & \nod & \nod & 31.67 & 31.57 & 31.43 & \nod & \nod & \nod & $31.61\pm 0.14$\\
V0857 & N4394 & U7523 & 12\phn 23\phn 24.7&$+$18\phn 29\phn 30 & 3 & A & 31.93 & \nod & \nod & \nod & 31.83 & \nod & \nod & \nod & \nod & $31.96\pm 0.08$\\
V0859 & \noc & U7522 & 12\phn 23\phn 25.8&$+$03\phn 42\phn 30 & 6 & W & \nod & \nod & \nod & \nod & \nod & 32.69 & \nod & 33.37 & \nod & $32.79\pm 0.07$\\
V0865 & N4396 & U7526 & 12\phn 23\phn 27.5&$+$15\phn 56\phn 55 & 7 & A & 30.58 & \nod & \nod & 30.62 & 30.25 & 30.33 & 31.00 & \nod & \nod & $30.58\pm 0.28$\\
V0873 & N4402 & U7528 & 12\phn 23\phn 34.8&$+$13\phn 23\phn 24 & 3 & A & 31.13 & \nod & \nod & 31.31 & 30.81 & 31.06 & 30.29 & 32.07 & \nod & $31.02\pm 0.35$\\
V0874 & N4405 & U7529 & 12\phn 23\phn 35.8&$+$16\phn 27\phn 26 & 0 & \nod & 31.52 & \nod & \nod & \nod & 30.15 & 30.63 & \nod & \nod & \nod & $30.80\pm 0.49$\\
V0905 & N4411A& U7537 & 12\phn 23\phn 56.4&$+$09\phn 08\phn 54 & 5 & B & 30.27 & \nod & \nod & \nod & 30.68 & 33.04 & \nod & \nod & \nod & $30.89$\tablenotemark{b}\\
V0912 & N4413 & U7538 & 12\phn 24\phn 00.0&$+$12\phn 53\phn 18 & 2 & A & 30.95 & \nod & \nod & 30.45 & 31.01 & 31.29 & 31.01 & \nod & \nod & $30.97\pm 0.33$\\
V0938 & N4416 & U7541 & 12\phn 24\phn 14.5&$+$08\phn 11\phn 51 & 6 & B & 33.12 & \nod & \nod & \nod & 32.36 & \nod & \nod & \nod & \nod & $32.82\pm 0.24$\\
V0939 & N4411B& U7546 & 12\phn 24\phn 15.0&$+$09\phn 09\phn 42 & 6 & B & 32.23 & \nod & \nod & \nod & 32.05 & \nod & \nod & \nod & \nod & $32.23\pm 0.05$\\
V0950 & I3356 & U7547 & 12\phn 24\phn 21.6&$+$11\phn 50\phn 18 & 10& A & 31.71 & \nod & 30.70 & \nod & 30.12 & \nod & \nod & \nod & \nod & $30.94\pm 0.55$\\
V0952 & \noc & \noc & 12\phn 24\phn 22.7&$+$10\phn 09\phn 17 & 5 & B & 32.48 & \nod & \nod & \nod & 31.74 & 31.55 & \nod & \nod & \nod & $31.97\pm 0.37$\\
V0957 & N4420 & U7549 & 12\phn 24\phn 24.6&$+$02\phn 46\phn 15 & 5 & W & \nod & \nod & \nod & \nod & \nod & 31.03 & 30.93 & \nod & \nod & $31.00\pm 0.01$\\
V0958 & N4419 & U7551 & 12\phn 24\phn 25.1&$+$15\phn 19\phn 28 & 1 & A & 30.73 & \nod & \nod & 30.82 & 30.15 & \nod & 30.97 & \nod & \nod & $30.71\pm 0.24$\\
V0971 & N4423 & U7556 & 12\phn 24\phn 36.2&$+$06\phn 09\phn 23 & 8 & \nod & 31.09 & \nod & \nod & \nod & 31.17 & 30.84 & 31.55 & \nod & \nod & $31.23\pm 0.34$\\
V0975 & \noc & U7557 & 12\phn 24\phn 36.0&$+$07\phn 32\phn 00 & 9 & B & 31.73 & \nod & \nod & \nod & 33.89 & \nod & \nod & \nod & \nod & \phm{000}\nod\tablenotemark{a}\\
V0979 & N4424 & U7561 & 12\phn 24\phn 40.2&$+$09\phn 41\phn 48 & 1 & B & 28.00 & \nod & \nod & 28.25 & \nod & \nod & \nod & \nod & \nod & $28.07\pm 0.09$\\
V0980 & I3365 & U7563 & 12\phn 24\phn 42.0&$+$16\phn 12\phn 00 & 10& A & 31.05 & \nod & \nod & 31.25 & 30.76 & 30.71 & \nod & \nod & \nod & $30.95\pm 0.19$\\
V0989 & \noc & \noc & 12\phn 24\phn 45.0&$+$07\phn 56\phn 54 & 7 & W' & 34.35 & \nod & \nod & \nod & \nod & \nod & \nod & \nod & \nod & $34.28        $\\
V0995 & I3371 & U7565 & 12\phn 24\phn 49.2&$+$11\phn 08\phn 36 & 6 & A & 32.16 & \nod & \nod & 31.99 & 31.95 & 31.33 & \nod & \nod & \nod & $31.87\pm 0.34$\\
V1002 & N4430 & U7566 & 12\phn 24\phn 53.6&$+$06\phn 32\phn 23 & 3 & B & 32.76 & \nod & \nod & \nod & 31.35 & \nod & \nod & \nod & \nod & $32.14\pm 0.57$\\
V1011 & \noc & U7567 & 12\phn 24\phn 56.7&$+$07\phn 55\phn 17 & 10& B & 32.53 & \nod & \nod & 32.31 & 31.70 & \nod & \nod & \nod & \nod & $32.21\pm 0.23$\\
V1043 & N4438 & U7574 & 12\phn 25\phn 13.8&$+$13\phn 17\phn 06 & 1 & A & \nod & 29.79 & \nod & \nod & 29.79 & 30.51 & 30.26 & 30.80 & \nod & $30.08\pm 0.27$\\
V1048 & \noc & U7579 & 12\phn 25\phn 22.6&$+$05\phn 59\phn 50 & 8 & W & 33.03 & \nod & \nod & \nod & 33.07 & 33.01 & \nod & \nod & \nod & $33.10\pm 0.14$\\
V1086 & N4445 & U7587 & 12\phn 25\phn 43.8&$+$09\phn 42\phn 48 & 2 & B & 31.54 & \nod & \nod & \nod & 31.29 & 31.30 & 31.26 & \nod & \nod & $31.41\pm 0.10$\\
V1091 & \noc & U7590 & 12\phn 25\phn 46.2&$+$09\phn 00\phn 18 & 4 & B & 31.79 & \nod & \nod & 32.03 & 31.67 & 31.25 & \nod & \nod & \nod & $31.70\pm 0.29$\\
V1110 & N4450 & U7594 & 12\phn 25\phn 58.0&$+$17\phn 21\phn 40 & 2 & A & 30.77 & 30.22 & 30.67 & 30.43 & 30.56 & 31.35 & 30.51 & 31.88 & \nod & $30.74\pm 0.37$\\
\noc & I3391 & U7595 & 12\phn 25\phn 55.6&$+$18\phn 41\phn 32 & 6 & \nod & 31.07 & \nod & \nod & \nod & 31.86 & \nod & \nod & \nod & \nod & $31.55\pm 0.52$\\
V1118 & N4451 & U7600 & 12\phn 26\phn 08.4&$+$09\phn 32\phn 06 & 3 & B & 32.23 & \nod & 31.97 & \nod & 32.04 & 32.37 & 32.16 & \nod & \nod & $32.29\pm 0.07$\\
V1126 & I3392 & U7602 & 12\phn 26\phn 12.0&$+$15\phn 16\phn 40 & 2 & A & 31.00 & \nod & \nod & \nod & 31.13 & 30.82 & 30.36 & \nod & \nod & $30.87\pm 0.35$\\
\noc & N4455 & U7603 & 12\phn 26\phn 14.1&$+$23\phn 06\phn 01 & 7 & \nod & 29.85 & \nod & \nod & 29.96 & \nod & 29.20 & \nod & \nod & \nod & $29.59\pm 0.36$\\
V1189 & I3414 & U7621 & 12\phn 26\phn 56.2&$+$07\phn 02\phn 50 & 8 & B & 31.24 & \nod & \nod & 31.01 & 30.58 & 31.01 & \nod & \nod & \nod & $30.97\pm 0.15$\\
V1193 & N4466 & U7626 & 12\phn 26\phn 58.0&$+$07\phn 58\phn 20 & 2 & B & 32.31 & \nod & \nod & 31.90 & 32.33 & 32.21 & \nod & \nod & \nod & $32.21\pm 0.26$\\
V1205 & N4470 & U7627 & 12\phn 27\phn 05.3&$+$08\phn 05\phn 56 & 1 & \nod & 30.82 & \nod & \nod & \nod & 30.77 & 31.13 & 31.26 & \nod & \nod & $31.05\pm 0.21$\\
V1290 & N4480 & U7647 & 12\phn 27\phn 53.4&$+$04\phn 31\phn 27 & 5 & W & \nod & \nod & \nod & \nod & \nod & 32.94 & 32.64 & 33.36 & \nod & $32.90\pm 0.06$\\
V1330 & N4492 & U7656 & 12\phn 28\phn 27.4&$+$08\phn 21\phn 13 & 1 & B & 31.69 & \nod & \nod & \nod & 32.61 & \nod & \nod & \nod & \nod & $32.23\pm 0.59$\\
V1356 & I3446 & \noc & 12\phn 28\phn 51.6&$+$11\phn 46\phn 00 & 9 & A & 33.48 & \nod & \nod & 32.62 & 31.91 & \nod & \nod & \nod & \nod & $32.69\pm 0.54$\\
V1375 & N4496A& U7668A& 12\phn 29\phn 05.8&$+$04\phn 12\phn 56 & 8 & W & \nod & \nod & \nod & \nod & \nod & \nod & \nod & \nod & 30.86 & $30.86\pm 0.03$\\
V1379 & N4498 & U7669 & 12\phn 29\phn 08.8&$+$17\phn 07\phn 46 & 6 & A & 30.82 & 30.34 & 30.36 & 30.63 & 30.72 & 30.98 & 30.99 & \nod & \nod & $30.72\pm 0.26$\\
V1393 & I0797 & U7676 & 12\phn 29\phn 22.9&$+$15\phn 24\phn 00 & 6 & \nod & 31.49 & \nod & \nod & \nod & 31.60 & 31.37 & 31.83 & \nod & \nod & $31.65\pm 0.27$\\
V1401 & N4501 & U7675 & 12\phn 29\phn 27.6&$+$14\phn 41\phn 42 & 3 & A & 31.20 & 31.12 & 31.25 & 31.13 & 30.82 & 31.58 & 30.87 & 31.98 & \nod & $31.22\pm 0.20$\\
V1410 & N4502 & U7677 & 12\phn 29\phn 32.2&$+$16\phn 57\phn 47 & 6 & A & 32.53 & \nod & \nod & 32.80 & 32.25 & \nod & \nod & \nod & \nod & $32.55\pm 0.11$\\
V1442 & I3474 & U7687 & 12\phn 30\phn 04.0&$+$02\phn 56\phn 18 & 7 & W & \nod & \nod & \nod & \nod & \nod & 30.95 & \nod & \nod & \nod & $30.90        $\\
V1450 & I3476 & U7695 & 12\phn 30\phn 10.8&$+$14\phn 19\phn 30 & 10& A & 30.90 & \nod & \nod & \nod & 31.28 & 31.71 & \nod & \nod & \nod & $31.35\pm 0.36$\\
\noc & \noc & U7697 & 12\phn 30\phn 20.8&$+$20\phn 27\phn 40 & 6 & \nod & \nod & \nod & \nod & \nod & \nod & 32.28 & \nod & \nod & \nod & $32.29        $\\
V1486 & I3483 & \noc & 12\phn 30\phn 39.0&$+$11\phn 37\phn 12 & 3 & A & 32.30 & \nod & \nod & \nod & 33.63 & 33.73 & \nod & \nod & \nod & $33.83\pm 0.03$\\
V1508 & N4519 & U7709 & 12\phn 30\phn 57.6&$+$08\phn 55\phn 48 & 7 & B & 31.27 & 31.82 & 31.25 & 31.28 & 31.38 & 31.84 & 31.33 & 32.57 & \nod & $31.61\pm 0.32$\\
V1516 & N4522 & U7711 & 12\phn 31\phn 07.8&$+$09\phn 27\phn 00 & 6 & B & 31.06 & 30.48 & 30.67 & 31.23 & 30.45 & 30.92 & 30.71 & \nod & \nod & $30.82\pm 0.20$\\
V1524 & N4523 & U7713 & 12\phn 31\phn 18.0&$+$15\phn 26\phn 00 & 9 & A & 32.16 & \nod & \nod & \nod & 33.25 & \nod & \nod & \nod & \nod & $32.79\pm 0.68$\\
V1532 & I0800 & U7716 & 12\phn 31\phn 25.8&$+$15\phn 37\phn 51 & 5 & \nod & 30.66 & \nod & \nod & \nod & 31.29 & 31.05 & 30.93 & \nod & \nod & $31.04\pm 0.32$\\
V1540 & N4527 & U7721 & 12\phn 31\phn 35.5&$+$02\phn 55\phn 45 & 4 & W & \nod & \nod & \nod & \nod & \nod & 31.21 & 30.03 & 31.38 & \nod & $30.62\pm 0.47$\\
V1554 & N4532 & U7726 & 12\phn 31\phn 46.7&$+$06\phn 44\phn 43 & 10& B & 29.94 & 31.10 & 30.65 & 31.01 & 30.01 & \nod & 30.23 & \nod & \nod & $30.55\pm 0.46$\\
V1555 & N4535 & U7727 & 12\phn 31\phn 47.9&$+$08\phn 28\phn 25 & 5 & B & 30.47 & 31.19 & 30.60 & 30.51 & 30.16 & 30.78 & 30.67 & 31.68 & 30.99 & $30.99\pm 0.05$\\
V1557 & N4533 & U7725 & 12\phn 31\phn 48.7&$+$02\phn 36\phn 10 & 7 & W & \nod & \nod & \nod & \nod & \nod & 31.43 & \nod & \nod & \nod & $31.40        $\\
V1562 & N4536 & U7732 & 12\phn 31\phn 53.5&$+$02\phn 27\phn 50 & 4 & W & \nod & \nod & \nod & \nod & \nod & 30.74 & 30.65 & \nod & 30.87 & $30.87\pm 0.04$\\
V1566 & I3517 & U7733 & 12\phn 31\phn 58.8&$+$09\phn 25\phn 54 & 9 & B & 31.96 & \nod & \nod & 31.70 & 31.26 & 32.11 & \nod & \nod & \nod & $31.78\pm 0.25$\\
V1569 & I3520 & \noc & 12\phn 32\phn 00.0&$+$13\phn 46\phn 54 & 6 & A & 33.47 & \nod & \nod & 31.65 & 32.03 & 31.73 & \nod & \nod & \nod & $31.84\pm 0.29$\\
V1575 & I3521 & U7736 & 12\phn 32\phn 06.8&$+$07\phn 26\phn 09 & 10& B & 31.28 & \nod & \nod & 31.79 & 31.07 & \nod & 31.23 & \nod & \nod & $31.39\pm 0.18$\\
V1581 & \noc & U7739 & 12\phn 32\phn 13.0&$+$06\phn 34\phn 35 & 10& B & 32.58 & \nod & \nod & \nod & 32.77 & \nod & \nod & \nod & \nod & $32.76\pm 0.23$\\
V1588 & N4540 & U7742 & 12\phn 32\phn 20.1&$+$15\phn 49\phn 37 & 6 & A & 31.83 & \nod & \nod & \nod & 30.39 & 31.39 & 30.51 & \nod & \nod & $31.07\pm 0.52$\\
V1605 & \noc & \noc & 12\phn 32\phn 42.5&$+$10\phn 42\phn 24 & 5 & A & 32.45 & \nod & \nod & \nod & 32.04 & 31.25 & \nod & \nod & \nod & $31.96\pm 0.53$\\
V1615 & N4548 & U7753 & 12\phn 32\phn 55.2&$+$14\phn 46\phn 24 & 3 & A & 30.81 & \nod & 30.89 & \nod & 30.66 & 31.31 & 30.88 & 31.65 & 31.05 & $31.05\pm 0.05$\\
V1624 & N4544 & U7756 & 12\phn 33\phn 03.3&$+$03\phn 18\phn 45 & 1 & S & \nod & \nod & \nod & 31.73 & \nod & 31.68 & 31.17 & \nod & \nod & $31.52\pm 0.18$\\
V1644 & \noc & \noc & 12\phn 33\phn 21.0&$+$14\phn 08\phn 00 & 9 & A & 33.17 & \nod & \nod & 33.28 & 32.75 & \nod & \nod & \nod & \nod & $33.09\pm 0.09$\\
\noc & N4561 & U7768 & 12\phn 33\phn 38.4&$+$19\phn 35\phn 56 & 7 & \nod & 32.01 & \nod & \nod & \nod & 31.42 & \nod & \nod & \nod & \nod & $31.80\pm 0.16$\\
V1673 & N4567 & U7777 & 12\phn 34\phn 01.2&$+$11\phn 32\phn 00 & 4 & A & \nod & \nod & \nod & \nod & \nod & \nod & \nod & 32.68 & \nod & $32.10        $\\
V1676 & N4568 & U7776 & 12\phn 34\phn 02.4&$+$11\phn 30\phn 54 & 4 & A & \nod & \nod & \nod & \nod & \nod & 31.36 & \nod & 32.10 & \nod & $31.39\pm 0.07$\\
V1678 & I3576 & U7781 & 12\phn 34\phn 06.0&$+$06\phn 54\phn 00 & 9 & B & 33.30 & \nod & \nod & \nod & 30.41 & \nod & \nod & \nod & \nod & \phm{000}\nod\tablenotemark{a}\\
V1686 & I3583 & U7784 & 12\phn 34\phn 12.6&$+$13\phn 32\phn 00 & 10& A & 30.17 & \nod & \nod & 31.17 & 30.46 & \nod & 30.68 & \nod & \nod & $30.66\pm 0.34$\\
V1690 & N4569 & U7786 & 12\phn 34\phn 18.6&$+$13\phn 26\phn 24 & 2 & A & 29.77 & 29.56 & \nod & 29.49 & 29.69 & 30.05 & 30.37 & 31.03 & \nod & $29.87\pm 0.37$\\
V1696 & N4571 & U7788 & 12\phn 34\phn 25.2&$+$14\phn 29\phn 48 & 6 & A & 30.99 & \nod & 30.55 & \nod & 30.84 & 31.70 & 31.41 & \nod & \nod & $31.18\pm 0.38$\\
V1699 & I3591 & U7790 & 12\phn 34\phn 29.9&$+$07\phn 11\phn 59 & 10& B & 31.02 & \nod & \nod & 31.20 & 30.69 & \nod & \nod & \nod & \nod & $31.00\pm 0.09$\\
V1725 & \noc & \noc & 12\phn 35\phn 09.0&$+$08\phn 50\phn 00 & 10& B & 31.66 & \nod & \nod & 31.57 & 30.71 & \nod & \nod & \nod & \nod & $31.34\pm 0.30$\\
V1726 & \noc & U7795 & 12\phn 35\phn 13.0&$+$07\phn 22\phn 47 & 10& B & 31.08 & \nod & \nod & \nod & 31.03 & 31.20 & \nod & \nod & \nod & $31.15\pm 0.09$\\
V1727 & N4579 & U7796 & 12\phn 35\phn 12.0&$+$12\phn 05\phn 36 & 3 & A & 31.30 & \nod & 31.06 & \nod & 31.07 & 31.74 & 31.61 & 31.86 & \nod & $31.41\pm 0.24$\\
V1730 & N4580 & U7794 & 12\phn 35\phn 15.6&$+$05\phn 38\phn 38 & 2 & \nod & 31.35 & \nod & \nod & \nod & 31.34 & \nod & \nod & \nod & \nod & $31.43\pm 0.12$\\
V1758 & \noc & U7802 & 12\phn 35\phn 48.0&$+$08\phn 10\phn 00 & 6 & \nod & 31.73 & \nod & \nod & \nod & 32.07 & 31.65 & \nod & \nod & \nod & $31.87\pm 0.30$\\
V1760 & N4586 & U7804 & 12\phn 35\phn 55.1&$+$04\phn 35\phn 37 & 1 & S & \nod & \nod & \nod & 31.24 & \nod & \nod & 30.40 & \nod & \nod & $30.79\pm 0.36$\\
V1780 & N4591 & U7821 & 12\phn 36\phn 39.9&$+$06\phn 17\phn 11 & 3 & \nod & 33.25 & \nod & \nod & \nod & \nod & 33.64 & \nod & 33.80 & \nod & $33.41\pm 0.22$\\
V1791 & I3617 & U7822 & 12\phn 36\phn 53.0&$+$08\phn 14\phn 12 & 10& \nod & 31.34 & \nod & 31.06 & 31.89 & 30.92 & \nod & \nod & \nod & \nod & $31.37\pm 0.26$\\
V1811 & N4595 & U7826 & 12\phn 37\phn 20.9&$+$15\phn 34\phn 23 & 3 & A & 30.66 & \nod & \nod & 30.91 & 30.78 & 30.71 & 30.94 & \nod & \nod & $30.82\pm 0.17$\\
V1859 & N4606 & U7839 & 12\phn 38\phn 26.4&$+$12\phn 11\phn 12 & 1 & A & 30.52 & \nod & \nod & 30.58 & 30.37 & \nod & \nod & \nod & \nod & $30.52\pm 0.04$\\
V1868 & N4607 & U7843 & 12\phn 38\phn 40.8&$+$12\phn 09\phn 54 & 3 & A & 31.53 & \nod & \nod & 31.86 & 31.36 & 31.49 & \nod & \nod & \nod & $31.58\pm 0.12$\\
V1923 & N4630 & U7871 & 12\phn 39\phn 58.5&$+$04\phn 14\phn 03 & 10& S & \nod & \nod & \nod & \nod & \nod & 31.14 & 30.91 & \nod & \nod & $31.05\pm 0.06$\\
V1929 & N4633 & U7874 & 12\phn 40\phn 06.6&$+$14\phn 37\phn 48 & 8 & A & 31.63 & \nod & \nod & 32.57 & 31.45 & 31.86 & 31.48 & \nod & \nod & $31.84\pm 0.33$\\
V1932 & N4634 & U7875 & 12\phn 40\phn 10.2&$+$14\phn 34\phn 12 & 6 & A & 31.53 & \nod & \nod & \nod & 31.59 & 31.27 & \nod & \nod & \nod & $31.51\pm 0.23$\\
V1933 & \noc & \noc & 12\phn 40\phn 12.5&$+$07\phn 36\phn 42 & 2 & \nod & 32.82 & \nod & \nod & \nod & 32.49 & \nod & \nod & \nod & \nod & $32.73\pm 0.02$\\
V1943 & N4639 & U7884 & 12\phn 40\phn 21.0&$+$13\phn 31\phn 54 & 4 & A & 31.74 & \nod & 31.57 & 32.07 & 31.74 & 32.55 & 32.00 & 32.53 & 31.71 & $31.71\pm 0.08$\\
V1955 & N4641 & U7889 & 12\phn 40\phn 36.0&$+$12\phn 19\phn 24 & -2& A & 29.64 & \nod & \nod & \nod & 29.17 & \nod & \nod & \nod & \nod & $29.50\pm 0.12$\\
V1972 & N4647 & U7896 & 12\phn 41\phn 01.2&$+$11\phn 51\phn 12 & 5 & A & \nod & \nod & 30.65 & \nod & 31.09 & 31.57 & 31.19 & 31.95 & \nod & $31.25\pm 0.26$\\
\noc & N4651 & U7901 & 12\phn 41\phn 12.5&$+$16\phn 40\phn 05 & 5 & \nod & 31.57 & 31.45 & 31.58 & 31.46 & 31.46 & 32.16 & \nod & 32.57 & \nod & $31.74\pm 0.25$\\
V1987 & N4654 & U7902 & 12\phn 41\phn 26.4&$+$13\phn 24\phn 00 & 6 & A & 30.56 & 30.79 & 30.43 & 30.74 & 30.51 & 30.74 & 30.75 & 31.41 & \nod & $30.69\pm 0.11$\\
V2023 & I3742 & U7932 & 12\phn 43\phn 01.2&$+$13\phn 36\phn 24 & 5 & A & 31.69 & \nod & \nod & 31.90 & 31.37 & 31.73 & 32.25 & \nod & \nod & $31.84\pm 0.32$\\
V2058 & N4689 & U7965 & 12\phn 45\phn 15.0&$+$14\phn 02\phn 06 & 5 & A & 30.75 & \nod & 30.64 & \nod & 30.64 & 31.06 & \nod & 31.74 & \nod & $30.88\pm 0.14$\\
V2070 & N4698 & U7970 & 12\phn 45\phn 51.8&$+$08\phn 45\phn 37 & 1 & \nod & 32.80 & 31.28 & 31.77 & 31.73 & 32.03 & \nod & 31.53 & 32.35 & \nod & $31.93\pm 0.41$\\
\noc & N4701 & U7975 & 12\phn 46\phn 39.0&$+$03\phn 39\phn 45 & 6 & \nod & \nod & \nod & \nod & \nod & \nod & \nod & \nod & 32.71 & \nod & $32.13        $\\
\noc & N4713 & U7985 & 12\phn 47\phn 25.6&$+$05\phn 34\phn 58 & 7 & \nod & 30.92 & \nod & \nod & 30.63 & \nod & 30.65 & \nod & \nod & \nod & $30.67\pm 0.15$\\
\noc & N4725 & U7989 & 12\phn 47\phn 59.9&$+$25\phn 46\phn 20 & 2 & \nod & \nod & \nod & \nod & \nod & \nod & \nod & \nod & 30.78 & 30.46 & $30.46\pm 0.06        $\\
\noc & N4746 & U8007 & 12\phn 49\phn 25.2&$+$12\phn 21\phn 18 & 3 & \nod & 32.43 & \nod & \nod & \nod & 32.34 & 32.59 & \nod & 33.15 & \nod & $32.54\pm 0.10$\\
\noc & N4758 & U8014 & 12\phn 50\phn 14.8&$+$16\phn 07\phn 10 & 9 & \nod & 30.57 & 30.62 & \nod & 31.19 & 30.95 & 30.93 & \nod & \nod & \nod & $30.87\pm 0.24$\\
\noc & N4771 & U8020 & 12\phn 50\phn 48.5&$+$01\phn 32\phn 30 & 6 & \nod & \nod & \nod & \nod & \nod & \nod & \nod & \nod & 32.31 & \nod & $31.69        $\\
\noc & N4772 & U8021 & 12\phn 50\phn 55.9&$+$02\phn 26\phn 27 & 1 & \nod & \nod & \nod & \nod & \nod & \nod & \nod & \nod & 32.81 & \nod & $32.24        $\\
\noc & I3881 & U8036 & 12\phn 52\phn 20.2&$+$19\phn 26\phn 55 & 6 & \nod & 31.90 & \nod & \nod & 31.08 & \nod & 31.11 & \nod & \nod & \nod & $31.30\pm 0.39$\\
\noc & N4808 & U8054 & 12\phn 53\phn 17.0&$+$04\phn 34\phn 28 & 6 & \nod & \nod & \nod & \nod & 31.49 & \nod & 30.80 & \nod & \nod & \nod & $31.07\pm 0.32$\\
\noc & N4845 & U8078 & 12\phn 55\phn 28.1&$+$01\phn 50\phn 42 & 2 & \nod & \nod & \nod & \nod & \nod & \nod & \nod & \nod & 31.81 & \nod & $31.15        $\\
\noc & \noc & U8085 & 12\phn 55\phn 47.8&$+$14\phn 49\phn 52 & 6 & \nod & \nod & \nod & \nod & \nod & \nod & 32.54 & \nod & \nod & \nod & $32.56        $\\
\noc & \noc & U8114 & 12\phn 57\phn 54.4&$+$13\phn 56\phn 35 & 3 & \nod & \nod & \nod & \nod & \nod & \nod & 32.14 & \nod & \nod & \nod & $32.14        $\\
\enddata
\vskip-0.5cm
\tablecomments{Col.\ (1)$-$(3): Galaxy names according to VCC, 
NGC or IC, and UGC or CGCG.
\newline\phn\phn
Col.\ (4) \& (5): Galaxy coordinates. Units of right ascension are
hours, minutes, and seconds. Units of declination are degrees,
arcminutes, and arcseconds.
\newline\phn\phn
Col.\ (6): Morphological type index defined in RC3.
\newline\phn\phn
Col.\ (7): Membership given in VCC and \citet{BPT93}. M:
member; A/B: clusters A/B; W/W'/M: clouds W/W'/M; S: southern
extension.
\newline\phn\phn
Col.\ (8)$-$(15): Distance modulus to each galaxy given in each of the
data sets listed in Table~\ref{datasets}.
\newline\phn\phn
Col.\ (16): Distance modulus from Cepheids given by \citet{Fre01}.
\newline\phn\phn
Col.\ (17): Mean distance modulus and $1\sigma$ uncertainty resulting from the homogeneization of the data sets (see text).
\newline\phn\phn
\tablenotemark{a}\ \ All individual measures of the distance modulus rejected in the homogenization procedure.
\newline\phn\phn
\tablenotemark{b}\ \ All but one individual measure of the distance modulus rejected in the homogenization procedure.
}
\end{deluxetable}



\clearpage



\begin{deluxetable}{lrlrrccr}
\tablewidth{386.39116pt}
\tablenum{3}
\tablecaption{Parameters for the Members of the 21-cm Sample\label{parameters}}
\tablecolumns{8}
\tablehead{ 
\multicolumn{1}{c}{Galaxy} & \multicolumn{1}{c}{\vel} & 
\multicolumn{1}{c}{d} & & & \multicolumn{1}{c}{\wc} & & \\
\multicolumn{1}{c}{Name} & \multicolumn{1}{c}{(\kms)} & \multicolumn{1}{c}{(Mpc)} &
\multicolumn{1}{c}{$\dfbis$} & \multicolumn{1}{c}{$\df$} & 
\multicolumn{1}{c}{(\kms)} & \multicolumn{1}{c}{Flag} &
\multicolumn{1}{c}{\mtc}\\
\multicolumn{1}{c}{(1)} & \multicolumn{1}{c}{(2)} & \multicolumn{1}{c}{(3)} &
\multicolumn{1}{c}{(4)} & \multicolumn{1}{c}{(5)} &
\multicolumn{1}{c}{(6)} & \multicolumn{1}{c}{(7)} &
\multicolumn{1}{c}{(8)}
}
\startdata
Z69-10& 1330\phs & $12.08^{+2.24}_{-1.89}$ & $-$0.19\phd & $-$0.35\phd & 107 & w	 & $-$15.83\\
I0755 & 1388\phs & $18.45^{+1.59}_{-1.47}$ &  0.21\phd &  0.09\phd & 231 & H	 & $-$17.69\\
N4037 &  814\phs & $14.00^{+0.59}_{-0.57}$ &  0.88\phd &  0.67\phd & 190 & i	 & $-$17.96\\
N4064 &  837\phs & $ 9.82^{+0.42}_{-0.40}$ &  1.39\phd &  1.19\phd & 226 & \noc  & $-$18.15\\
N4067 & 2281\phs & $43.65^{+4.21}_{-3.77}$ & $-$0.08\phd & $-$0.19\phd & 422 & \noc  & $-$20.56\\
U7133 & 2486\phs & $65.16		 $ & $-$0.05\phd &  0.07\phd & 260 & o,W   & $-$19.26\\
V0015 & 2406\phs & $55.46		 $ &  0.77\phd &  0.88\phd & 227 & o	 & $-$19.45\\
V0025 & 2062\phs & $31.92^{+5.76}_{-4.80}$ & $-$0.45\phd & $-$0.47\phd & 379 & i,s	 & $-$19.91\\
V0034 &  145\phs & $25.47^{+0.96}_{-0.89}$ &  0.04\phd &  0.00\phd & 159 & \noc  & $-$17.46\\
V0047 & 1742\phs & $55.72		 $ &  0.40\phd &  0.38\phd & 342 & o	 & $-$19.74\\
V0058 & 2110\phs & $38.55^{+5.71}_{-4.91}$ & $-$0.19\phd & $-$0.04\phd & 323 & s         & $-$20.00\\
V0066 &  242\phs & $13.12^{+2.65}_{-2.24}$ &  0.02\phd &  0.08\phd & 296 & \noc  & $-$19.54\\
V0067 & $-$299\phs & $16.22^{+2.75}_{-2.31}$ &  0.23\phd &  0.20\phd & 171 & \noc  & $-$17.41\\
V0073 & 1933\phs & $42.27^{+6.04}_{-5.21}$ &  0.05\phd &  0.06\phd & 460 & s     & $-$20.43\\
V0081 & 1963\phs & $47.21^{+13.61}_{-10.46}$& 0.01\phd &  0.10\phd & 209 & i,s	 & $-$17.80\\
V0087 & $-$242\phs & $24.21^{+0.45}_{-0.48}$ &  0.36\phd &  0.36\phd & 151 & \noc  & $-$16.86\\
V0089 & 2004\phs & $30.06^{+4.61}_{-4.04}$ &  0.20\phd &  0.28\phd & 438 & \noc  & $-$20.32\\
V0092 & $-$246\phs & $13.55^{+1.24}_{-1.14}$ &  0.26\phd &  0.45\phd & 479 & \noc  & $-$20.61\\
V0097 & 2356\phs & $38.19^{+2.36}_{-2.31}$ &  0.20\phd &  0.32\phd & 388 & \noc  & $-$20.35\\
V0120 & 1910\phs & $23.12^{+2.00}_{-1.82}$ & $-$0.01\phd &  0.07\phd & 295 & \noc  & $-$19.52\\
V0131 & 2201\phs & $38.02^{+2.16}_{-2.09}$ &  0.05\phd &  0.06\phd & 315 & \noc  & $-$19.86\\
V0143 &  264\phs & $31.19^{+5.12}_{-4.45}$ &  0.54\phd &  0.30\phd & 177 & H,s	 & $-$17.74\\
V0145 &  584\phs & $18.71^{+1.34}_{-1.22}$ &  0.16\phd &  0.34\phd & 295 & \noc  & $-$19.76\\
V0152 &  456\phs & $21.88^{+2.90}_{-2.60}$ &  0.00\phd & $-$0.02\phd & 267 & \noc  & $-$18.71\\
V0157 & $-$200\phs & $18.71^{+1.52}_{-1.44}$ &  0.17\phd &  0.12\phd & 368 & \noc  & $-$19.98\\
V0162 & 1847\phs & $31.05^{+4.27}_{-3.78}$ & $-$0.22\phd & $-$0.16\phd & 238 & \noc  & $-$18.17\\
V0167 &   16\phs & $15.56^{+2.22}_{-1.94}$ &  0.53\phd &  0.71\phd & 533 & \noc  & $-$21.02\\
V0187 &  114\phs & $23.44^{+3.47}_{-3.00}$ &  0.02\phd &  0.09\phd & 248 & \noc  & $-$19.46\\
V0199 & 2448\phs & $44.46		 $ &  0.25\phd &  0.33\phd & 575 & o	 & $-$20.74\\
V0213 & $-$278\phs & $40.93^{+5.00}_{-4.54}$ & $-$0.12\phd & $-$0.25\phd & 308 & i,s	 & $-$18.98\\
V0222 & 2453\phs & $17.95^{+1.55}_{-1.39}$ &  0.35\phd &  0.34\phd & 343 & H	 & $-$19.23\\
V0224 & 2011\phs & $31.62^{+2.89}_{-2.60}$ &  0.22\phd &  0.28\phd & 227 & \noc  & $-$18.89\\
V0226 &  756\phs & $23.55^{+2.15}_{-1.95}$ &  0.54\phd &  0.47\phd & 330 & \noc  & $-$19.81\\
V0267 &  584\phs & $27.93^{+0.65}_{-0.64}$ & $-$0.01\phd &  0.00\phd & 644 & \noc  & $-$18.50\\
V0289 &  709\phs & $21.68^{+0.81}_{-0.79}$ &  0.37\phd &  0.04\phd & 180 & \noc  & $-$17.37\\
V0297 & 1847\phs & $23.33		 $ &  0.54\phd &  0.43\phd & 170 & o,W   & $-$17.04\\
V0307 & 2303\phs & $16.14^{+3.35}_{-2.76}$ & $-$0.04\phd & $-$0.04\phd & 452 & i	 & $-$20.90\\
V0318 & 2328\phs & $29.79^{+2.43}_{-2.25}$ & $-$0.09\phd & $-$0.05\phd & 236 & \noc  & $-$18.28\\
V0341 & 1649\phs & $39.63^{+1.68}_{-1.70}$ & $-$0.03\phd &  0.04\phd & 536 & H   & $-$20.63\\
V0382 & 2219\phs & $27.54^{+3.36}_{-2.99}$ &  0.32\phd &  0.29\phd & 365 & \noc  & $-$20.29\\
V0393 & 2473\phs & $27.29		 $ &  0.13\phd &  0.14\phd & 273 & i,o   & $-$18.54\\
V0404 & 1574\phs & $34.67		 $ &  0.32\phd &  0.38\phd & 268 & o,W   & $-$19.24\\
V0415 & 2413\phs & $30.62^{+14.05}_{-9.63}$& 0.49\phd &  0.41\phd & 200 & s     & $-$17.08\\
V0449 & 2377\phs & $44.87^{+0.21}_{-0.20}$ &  0.06\phd &  0.23\phd & 371 & W	 & $-$20.47\\
V0453 &  785\phs & $27.29^{+6.75}_{-5.41}$ &  0.56\phd &  0.51\phd & 127 & s,w	 & $-$16.14\\
V0465 &  239\phs & $14.66^{+1.71}_{-1.54}$ & $-$0.08\phd & $-$0.08\phd & 244 & \noc  & $-$19.04\\
V0483 & 1016\phs & $18.88^{+4.24}_{-3.48}$ &  0.08\phd &  0.03\phd & 318 & \noc  & $-$19.92\\
V0491 &  112\phs & $34.83		 $ & $-$0.40\phd & $-$0.36\phd & 349 & i,o,W & $-$19.91\\
V0497 & 1034\phs & $19.41^{+3.19}_{-2.72}$ &  0.27\phd &  0.28\phd & 383 & \noc  & $-$20.39\\
V0509 & 1109\phs & $34.83^{+4.61}_{-4.11}$ &  0.03\phd &  0.04\phd & 181 & \noc  & $-$17.53\\
V0522 & 1814\phs & $40.74		 $ &  0.92\phd &  0.95\phd & 457 & H,o,W & $-$20.13\\
V0524 &  913\phs & $27.42^{+3.49}_{-3.07}$ &  1.16\phd &  1.24\phd & 490 & W	 & $-$20.33\\
V0559 &   47\phs & $10.76^{+1.31}_{-1.19}$ &  1.19\phd &  1.02\phd & 232 & \noc  & $-$18.39\\
V0566 & 1267\phs & $25.82^{+1.34}_{-1.27}$ & $-$0.14\phd & $-$0.20\phd & 117 & w	 & $-$16.79\\
V0567 & 2215\phs & $29.38^{+0.82}_{-0.86}$ &  0.11\phd &  0.13\phd & 233 & \noc  & $-$18.12\\
V0570 & 1319\phs & $14.00^{+1.49}_{-1.33}$ &  1.04\phd &  0.90\phd & 280 & \noc  & $-$18.94\\
V0576 & 1117\phs & $28.84^{+3.08}_{-2.80}$ &  0.19\phd &  0.26\phd & 334 & \noc  & $-$19.84\\
V0593 & 1392\phs & $23.55^{+2.87}_{-2.52}$ &  0.47\phd &  0.17\phd & 159 & H	 & $-$17.31\\
V0596 & 1468\phs & $15.21^{+0.50}_{-0.49}\tablenotemark{a}$ &  0.32\phd &  0.49\phd & 482 & i	  & $-$21.12\\
V0620 &  622\phs & $20.14^{+6.53}_{-4.90}$ & $-$0.13\phd & $-$0.28\phd & 130 & s,w	 & $-$16.40\\
V0630 & 1438\phs & $19.32^{+1.57}_{-1.44}$ &  0.65\phd &  0.73\phd & 279 & \noc  & $-$19.35\\
V0656 &  868\phs & $28.44^{+4.52}_{-3.92}$ &  0.40\phd &  0.40\phd & 401 & \noc  & $-$19.92\\
V0667 & 1279\phs & $25.59^{+0.96}_{-0.89}$ &  0.53\phd &  0.54\phd & 212 & \noc  & $-$18.39\\
V0692 & 2203\phs & $11.17^{+1.25}_{-1.13}$ &  0.60\phd &  0.17\phd & 164 & i	 & $-$17.47\\
V0697 & 1085\phs & $36.81		 $ &  0.30\phd &  0.29\phd & 280 & i,o   & $-$18.68\\
V0699 &  579\phs & $32.21^{+3.27}_{-2.97}$ & $-$0.31\phd & $-$0.46\phd & 295 & i	 & $-$18.44\\
V0713 &  998\phs & $29.38^{+3.74}_{-3.25}$ &  1.28\phd &  1.35\phd & 302 & \noc  & $-$19.42\\
V0737 & 1566\phs & $27.16		 $ &  0.02\phd &  0.00\phd & 164 & o,W   & $-$17.95\\
V0768 & 2354\phs & $33.57^{+1.10}_{-1.12}$ &  0.61\phd &  0.49\phd & 212 & H	 & $-$18.71\\
V0785 & 2395\phs & $47.86		 $ & $-$0.17\phd &  0.11\phd & 624 & i,o   & $-$21.00\\
V0792 &  840\phs & $21.88^{+4.06}_{-3.47}$ &  0.60\phd &  0.62\phd & 346 & \noc  & $-$19.51\\
V0801 & 1608\phs & $16.67^{+1.44}_{-1.32}$ & $-$0.42\phd & $-$0.73\phd & 302 & \noc  & $-$18.84\\
V0809 & $-$262\phs & $19.95^{+1.13}_{-1.05}$ &  0.62\phd &  0.61\phd & 178 & \noc  & $-$17.89\\
V0827 &  847\phs & $23.88^{+2.67}_{-2.36}$ &  0.07\phd &  0.15\phd & 306 & \noc  & $-$19.53\\
V0836 & 2420\phs & $16.60^{+1.52}_{-1.36}$ &  1.14\phd &  1.24\phd & 401 & \noc  & $-$20.17\\
V0849 &  972\phs & $23.12^{+4.68}_{-3.86}$ &  0.06\phd & $-$0.02\phd & 236 & i   & $-$18.73\\
V0851 & 1061\phs & $20.99^{+1.40}_{-1.31}$ &  0.46\phd &  0.48\phd & 216 & \noc  & $-$18.59\\
V0857 &  823\phs & $24.66^{+0.93}_{-0.95}$ &  0.27\phd &  0.33\phd & 432 & i	 & $-$20.41\\
V0859 & 1265\phs & $36.14^{+1.18}_{-1.15}$ &  0.33\phd &  0.43\phd & 321 & W	 & $-$19.94\\
V0865 & $-$228\phs & $13.06^{+1.80}_{-1.57}$ &  0.22\phd &  0.22\phd & 202 & \noc  & $-$18.54\\
V0873 &  121\phs & $16.00^{+2.80}_{-2.35}$ &  0.57\phd &  0.50\phd & 298 & \noc  & $-$19.27\\
V0905 & 1146\phs & $15.07		 $ &  0.52\phd &  0.41\phd & 147 & i,w   & $-$17.43\\
V0938 & 1244\phs & $36.64^{+4.28}_{-3.83}$ &  0.15\phd &  0.20\phd & 417 & i	 & $-$19.80\\
V0939 & 1135\phs & $27.93^{+0.65}_{-0.57}$ &  0.53\phd &  0.63\phd & 328 & i	 & $-$19.16\\
V0957 & 1530\phs & $15.85^{+0.07}_{-0.07}$ &  0.07\phd & $-$0.04\phd & 232 & W	 & $-$18.76\\
V0958 & $-$380\phs & $13.87^{+1.62}_{-1.43}$ &  0.92\phd &  0.74\phd & 307 & \noc  & $-$19.19\\
V0971 &  942\phs & $17.62^{+2.99}_{-2.55}$ & $-$0.08\phd & $-$0.10\phd & 191 & \noc  & $-$18.45\\
V0979 &  314\phs & $ 4.11^{+0.17}_{-0.18}$ &  1.32\phd &  0.76\phd & 108 & w	 & $-$15.99\\
V0989 & 1704\phs & $71.78		 $ &  0.12\phd &  0.13\phd & 195 & i,o   & $-$18.52\\
V0995 &  797\phs & $23.66^{+4.01}_{-3.43}$ & $-$0.06\phd & $-$0.06\phd & 172 & \noc  & $-$17.57\\
V1002 & 1301\phs & $26.79^{+8.04}_{-6.12}$&  0.27\phd &  0.29\phd & 420 & i,s	 & $-$19.48\\
V1043 &  $-$45\phs & $10.38^{+1.37}_{-1.23}$ &  0.73\phd &  0.84\phd & 398 & W	 & $-$19.56\\
V1048 & 2102\phs & $41.69^{+2.78}_{-2.67}$ &  0.23\phd &  0.27\phd & 251 & \noc  & $-$19.14\\
V1086 &  230\phs & $19.14^{+0.90}_{-0.88}$ &  1.11\phd &  0.98\phd & 240 & \noc  & $-$18.61\\
V1091 &  983\phs & $21.88^{+3.13}_{-2.71}$ & $-$0.16\phd & $-$0.39\phd & 203 & \noc  & $-$18.06\\
V1110 & 1861\phs & $14.06^{+2.61}_{-2.18}$ &  1.06\phd &  1.08\phd & 405 & \noc  & $-$20.14\\
I3391 & 1618\phs & $20.42^{+5.52}_{-4.39}$ &  0.29\phd &  0.22\phd & 183 & i,s	 & $-$18.03\\
V1118 &  725\phs & $28.71^{+0.94}_{-0.85}$ &  0.43\phd &  0.25\phd & 308 & \noc  & $-$19.35\\
V1126 & 1571\phs & $14.93^{+2.61}_{-2.20}$ &  1.08\phd &  0.79\phd & 228 & \noc  & $-$18.44\\
N4455 &  576\phs & $ 8.28^{+1.49}_{-1.27}$ & $-$0.12\phd & $-$0.20\phd & 160 & \noc  & $-$16.77\\
V1189 &  381\phs & $15.63^{+1.12}_{-1.04}$ &  0.24\phd &  0.18\phd & 170 & \noc  & $-$17.31\\
V1193 &  616\phs & $27.67^{+3.52}_{-3.12}$ &  0.15\phd & $-$0.12\phd & 237 & \noc  & $-$18.24\\
V1205 & 2200\phs & $16.22^{+1.65}_{-1.54}$ &  0.07\phd & $-$0.34\phd & 206 & \noc  & $-$18.34\\
V1290 & 2281\phs & $38.02^{+1.07}_{-1.10}$ & $-$0.04\phd & $-$0.03\phd & 366 & W	 & $-$20.28\\
V1330 & 1638\phs & $27.93^{+8.72}_{-6.71}$&  0.85\phd &  0.78\phd & 297 & i,s	 & $-$19.19\\
V1356 & 1129\phs & $34.51^{+9.74}_{-7.65}$& $-$0.27\phd & $-$0.32\phd & 228 & s     & $-$17.94\\
V1375 & 1581\phs & $14.86^{+0.21}_{-0.20}$\tablenotemark{a} & $-$0.11\phd & $-$0.09\phd & 303 & W    & $-$19.09\\
V1379 & 1411\phs & $13.93^{+1.77}_{-1.57}$ &  0.22\phd &  0.23\phd & 234 & \noc  & $-$18.58\\
V1393 & 1995\phs & $21.38^{+2.83}_{-2.52}$ &  0.06\phd &  0.01\phd & 219 & \noc  & $-$18.37\\
V1401 & 2167\phs & $17.54^{+1.69}_{-1.56}$ &  0.40\phd &  0.55\phd & 603 & \noc  & $-$21.55\\
V1410 & 1531\phs & $32.36^{+1.68}_{-1.65}$ &  0.04\phd &  0.04\phd & 225 & \noc  & $-$18.58\\
V1442 & 1572\phs & $15.14		 $ & $-$0.06\phd & $-$0.09\phd & 154 & H,o,W & $-$17.71\\
U7697 & 2454\phs & $28.71		 $ &  0.09\phd &  0.13\phd & 215 & o,W   & $-$17.10\\
V1486 &    6\phs & $58.34^{+0.81}_{-0.67}$ &  0.65\phd &  0.65\phd & 180 & H	 & $-$18.82\\
V1508 & 1093\phs & $20.99^{+3.33}_{-2.83}$ & $-$0.21\phd & $-$0.14\phd & 287 & i	 & $-$19.37\\
V1516 & 2199\phs & $14.59^{+1.41}_{-1.30}$ &  0.65\phd &  0.69\phd & 257 & \noc  & $-$18.82\\
V1524 &  158\phs & $36.14^{+13.29}_{-9.72}$&$-$0.13\phd & $-$0.04\phd & 299 & i,s	 & $-$18.98\\
V1532 & 2237\phs & $16.14^{+2.56}_{-2.20}$ &  0.72\phd &  0.57\phd & 151 & i	 & $-$17.29\\
V1540 & 1571\phs & $13.30^{+3.22}_{-2.60}$ & $-$0.14\phd & $-$0.04\phd & 410 & W	 & $-$19.99\\
V1555 & 1825\phs & $15.78^{+0.37}_{-0.36}$\tablenotemark{a} &  0.16\phd &  0.19\phd & 397 & i	  & $-$20.64\\
V1557 & 1596\phs & $19.05		 $ &  0.34\phd &  0.35\phd & 187 & o,W   & $-$18.18\\
V1562 & 1641\phs & $14.93^{+0.28}_{-0.27}$\tablenotemark{a} &  0.08\phd &  0.25\phd & 394 & W	  & $-$20.43\\
V1566 &  295\phs & $22.70^{+2.77}_{-2.44}$ &  0.61\phd &  0.58\phd & 133 & w	 & $-$17.04\\
V1569 &  687\phs & $23.33^{+3.33}_{-2.95}$ &  0.90\phd &  0.85\phd & 215 & \noc  & $-$16.50\\
V1588 & 1184\phs & $16.37^{+4.43}_{-3.51}$ &  0.39\phd &  0.36\phd & 338 & i	 & $-$18.82\\
V1615 &  383\phs & $16.22^{+0.38}_{-0.37}$\tablenotemark{a} &  0.77\phd &  0.83\phd & 410 & i	  & $-$20.38\\
V1624 &  992\phs & $20.14^{+1.74}_{-1.63}$ &  0.57\phd &  0.43\phd & 212 & W	 & $-$18.01\\
V1644 &  646\phs & $41.50^{+1.76}_{-1.62}$ &  0.23\phd &  0.26\phd & 114 & w	 & $-$16.26\\
N4561 & 1324\phs & $22.91^{+1.75}_{-1.63}$ & $-$0.67\phd & $-$0.70\phd & 293 & \noc  & $-$19.23\\
V1676 & 2125\phs & $18.97^{+0.62}_{-0.64}$ &  0.35\phd &  0.49\phd & 248 & W	 & $-$20.43\\
V1690 & $-$328\phs & $ 9.42^{+1.75}_{-1.47}$ &  1.13\phd &  1.26\phd & 401 & \noc  & $-$20.28\\
V1696 &  235\phs & $17.22^{+3.29}_{-2.75}$ &  0.48\phd &  0.55\phd & 311 & i	 & $-$19.58\\
V1727 & 1397\phs & $19.14^{+2.24}_{-2.02}$ &  0.75\phd &  0.89\phd & 594 & i	 & $-$21.18\\
V1730 &  893\phs & $19.32^{+1.10}_{-1.08}$ &  1.17\phd &  1.05\phd & 283 & i	 & $-$18.84\\
V1758 & 1651\phs & $23.66^{+3.51}_{-3.01}$ &  0.37\phd &  0.37\phd & 185 & \noc  & $-$18.03\\
V1760 &  639\phs & $14.39^{+2.59}_{-2.20}$ &  1.15\phd &  1.06\phd & 292 & W	 & $-$18.72\\
V1780 & 2279\phs & $48.08^{+5.13}_{-4.63}$ &  0.17\phd &  0.23\phd & 399 & s     & $-$19.88\\
V1811 &  530\phs & $14.59^{+1.19}_{-1.10}$ &  0.34\phd & $-$0.02\phd & 204 & \noc  & $-$18.37\\
V1859 & 1528\phs & $12.71^{+0.24}_{-0.24}$ &  1.25\phd &  1.00\phd & 224 & \noc  & $-$18.27\\
V1868 & 2138\phs & $20.70^{+1.18}_{-1.09}$ &  0.73\phd &  0.69\phd & 245 & \noc  & $-$18.89\\
V1929 &  185\phs & $23.33^{+3.83}_{-3.27}$ &  0.06\phd &  0.07\phd & 228 & \noc  & $-$18.83\\
V1932 &   11\phs & $20.04^{+2.24}_{-2.05}$ &  0.28\phd &  0.30\phd & 291 & \noc  & $-$19.51\\
V1943 &  873\phs & $21.98^{+0.82}_{-0.80}$\tablenotemark{a} &  0.09\phd &  0.12\phd & 375 & \noc  & $-$19.90\\
V1972 & 1297\phs & $17.78^{+2.26}_{-1.97}$ &  0.42\phd &  0.36\phd & 345 & i,W   & $-$19.58\\
N4651 &  705\phs & $22.28^{+2.72}_{-2.45}$ & $-$0.29\phd & $-$0.29\phd & 472 & \noc  & $-$20.81\\
V1987 &  926\phs & $13.74^{+0.71}_{-0.67}$ &  0.06\phd &  0.12\phd & 365 & \noc  & $-$20.13\\
V2023 &  856\phs & $23.33^{+3.71}_{-3.24}$ & $-$0.17\phd & $-$0.26\phd & 205 & \noc  & $-$18.44\\
V2058 & 1493\phs & $15.00^{+1.00}_{-0.95}$ &  0.65\phd &  0.61\phd & 320 & i	 & $-$19.55\\
V2070 &  882\phs & $24.32^{+5.05}_{-4.23}$ & $-$0.20\phd & $-$0.14\phd & 663 & s     & $-$20.79\\
N4701 &  574\phs & $26.67		 $ & $-$0.30\phd & $-$0.21\phd & 272 & i,o,W & $-$19.55\\
N4713 &  511\phs & $13.61^{+0.97}_{-0.90}$ & $-$0.35\phd & $-$0.35\phd & 280 & i     & $-$18.66\\
N4725 & 1163\phs & $12.36^{+0.35}_{-0.34}$ &  0.21\phd &  0.45\phd & 510 & W     & $-$20.68\\
N4746 & 1669\phs & $32.21^{+1.52}_{-1.49}$ & $-$0.18\phd & $-$0.19\phd & 365 & \noc  & $-$20.08\\
N4758 & 1150\phs & $14.93^{+1.74}_{-1.56}$ &  0.21\phd &  0.21\phd & 207 & \noc  & $-$18.60\\
N4771 &  959\phs & $21.78		 $ &  0.31\phd &  0.39\phd & 293 & o,W   & $-$19.88\\
N4772 &  883\phs & $28.05		 $ & $-$0.07\phd & $-$0.03\phd & 510 & o,W   & $-$20.56\\
I3881 &  849\phs & $18.20^{+3.58}_{-3.01}$ &  0.08\phd &  0.14\phd & 249 & \noc  & $-$18.77\\
N4808 &  616\phs & $16.37^{+2.60}_{-2.28}$ & $-$0.68\phd & $-$0.69\phd & 265 & W	 & $-$19.60\\
N4845 &  940\phs & $16.98		 $ &  1.20\phd &  1.26\phd & 596 & o,W   & $-$19.90\\
U8085 & 1943\phs & $32.51		 $ & $-$0.14\phd & $-$0.06\phd & 246 & o,W   & $-$19.16\\
U8114 & 1887\phs & $26.79		 $ &  0.11\phd & $-$0.06\phd & 157 & o,W   & $-$16.90\\
\enddata
\tablecomments{Col.\ (2): Radial velocities from \hi\ measurements referred to the kinematic frame of the Local Group. 
\newline\phn\phn
Col.\ (3): Distance and $1\sigma$-errors.
\newline\phn\phn \tablenotemark{a}\ \ Distance from Cepheids. Errors
are those quoted by \citet{Fre01}.
\newline\phn\phn
Col.\ (4) \& (5): \hi\ deficiency parameters defined in the text.
\newline\phn\phn
Col.\ (6): Edge-on \hi\ line width.
\newline\phn\phn Col.\ (7): H: non-AGC flux or non-detection corrected only for
internal \hi\ self-absorption; i: $i < 45\degr$ (from LEDA); o: distance
based on a single non-Cepheid measurement; s: $1\sigma$ uncertainties
in the distance larger than 5 Mpc; W: corrected \hi\ line width \emph{not} from
\citeauthor{YFO97}; w: 100 $<$ \wc\ $< 150$ \kms\ (see text for further
details).
\newline\phn\phn
Col.\ (8): Total absolute corrected \bb-magnitude calculated from the absorption-free brightness given in LEDA.
}
\end{deluxetable}

\end{document}